\documentclass[%
 aip,
 amsmath,amssymb
 reprint,%
]{revtex4-1}
\usepackage{graphicx}
\usepackage{dcolumn}
\usepackage{bm}
\usepackage{amssymb}
\usepackage[dvipsnames]{xcolor}
\usepackage{soul}
\usepackage[utf8]{inputenc}
\usepackage[T1]{fontenc}
\usepackage{array}
\usepackage{mathptmx}
\usepackage{etoolbox} \usepackage{float}
\usepackage{multirow} \usepackage{tabularx} 
\usepackage{subcaption} 

\begin{document}


\title{Weakly Nonlinear Analysis of Phototactic Bioconvection under Thermal Buoyancy}
\author{Sandeep Kumar}
\email{sandeepkumar1.00123@gmail.com}
\affiliation{
Fluid Flow Systems Simulation Lab, Department of Energy Science and Engineering, Indian Institute of Technology Bombay, Mumbai 400076, India
}%
\author{Suneet Singh}
\affiliation{
Fluid Flow Systems Simulation Lab, Department of Energy Science and Engineering, Indian Institute of Technology Bombay, Mumbai 400076, India
}


\date{\today}

\begin{abstract}

A weakly nonlinear stability analysis is performed to investigate the influence of thermal buoyancy on phototactic bioconvection in a suspension of swimming microorganisms confined between horizontal boundaries and subjected to bottom heating or cooling. The objective is to examine how thermal forcing modifies the post-critical evolution of bioconvective instability through its influence on bifurcation behaviour and finite-amplitude convection. Linear stability theory is first employed to determine the critical conditions for the onset of convection. The corresponding direct and adjoint eigenvalue problems, together with the second-order resolvent system, are then solved numerically to derive the Stuart–Landau amplitude equation governing the nonlinear evolution of disturbances. A systematic analysis is carried out for varying thermal Rayleigh numbers. The results show that, as the thermal Rayleigh number approaches the classical Rayleigh–B\`{e}nard critical value, the critical bioconvection Rayleigh number decreases to zero, indicating a gradual transition from phototaxis-dominated to thermally driven convection. The computed positive Landau coefficient predicts supercritical pitchfork and Hopf bifurcations for stationary and oscillatory instabilities, respectively. In parameter regimes with oscillatory onset, increasing the thermal Rayleigh number drives a transition from a supercritical Hopf bifurcation to a supercritical pitchfork bifurcation. The nonlinear dynamics are further characterized through amplitude evolution, phase portraits, potential functions, and finite-amplitude flow structures.

\end{abstract}

\maketitle

\section{Introduction}

Bioconvection refers to the spontaneous formation of large-scale patterns in suspensions of motile microorganisms. The phenomenon was first observed experimentally by \citet{ref2}, while the term ``bioconvection'' was later introduced by \citet{ref3}. These microorganisms are slightly denser than the surrounding fluid. Under appropriate environmental conditions, their directed swimming leads to cell accumulation in the upper regions of the suspension, giving rise to an unstable density stratification. Experiments have shown that pattern formation ceases when the cells become immotile, emphasizing the crucial role of active swimming in bioconvective processes. However, bioconvection is not governed solely by density variations or vertical migration \citep{ref1}. Instead, microorganisms respond to environmental stimuli through directed movement, known as taxis. Common forms of taxis include gravitaxis, gyrotaxis, chemotaxis, and phototaxis. In phototaxis, cells move either toward a light source (positive phototaxis) or away from it (negative phototaxis). The present study focuses exclusively on phototactic behaviour. Experimental evidence indicates that both the intensity and nature of illumination significantly influence the structure and evolution of bioconvective patterns \citep{ref2,ref5}. Changes in light intensity can modify the size, shape, and spatial distribution of these patterns \citep{ref7}. This behaviour arises because microorganisms swim toward the light source when the total light intensity $K$ is below a critical value $K_c$, whereas they move away from it when this threshold is exceeded \citep{ref8}. In addition, the cells absorb light as they swim, altering the  light field and consequently affecting the resulting bioconvective structures.

The theoretical basis of phototactic bioconvection was established by \citet{ref9}, who proposed a continuum model for an absorbing, non-scattering suspension and investigated its hydrodynamic stability under uniform top illumination. This framework was later extended to two dimensions by \citet{ref13}, who examined the stability characteristics using a conservative finite-difference approach. Since both studies neglected light scattering, \citet{ref14} incorporated isotropic scattering into the model, and showed that scattering can lead to the formation of two distinct horizontal cell layers under suitable conditions. Subsequently, \citet{kumar2023} reported that suspensions bounded by a rigid upper surface are more stable than those with a stress-free boundary. Further developments include the study of forward anisotropic scattering by \citet{ref15} and the investigations of collimated and diffuse illumination by Panda et al. \citep{ref16, panda2020effects}. The effect of oblique irradiation was examined for non-scattering suspensions by \citet{ref17,panda2023phototactic} and later for isotropically scattering suspensions by \citet{ref18}, revealing significant changes in stability and cell distribution. The influence of rotation on phototactic bioconvection was first explored by \citet{kumar2023effect}, who demonstrated its stabilizing effect on non-scattering suspensions. Recently, \citet{kumar2024heating, kumar2024thermal} investigated the combined effects of phototaxis and thermal gradients for heating from below \citep{kumar2024thermal} and above \citep{kumar2024heating}, providing valuable insights into the resulting stability behaviour and flow dynamics. 



Despite considerable progress, most investigations of phototactic bioconvection have been confined to linear stability analysis, which determines the critical conditions for the onset of convection. However, such analyses do not describe the subsequent nonlinear evolution of the flow or the associated bifurcation behaviour. This limitation is particularly important in phototactic bioconvection, where microorganism concentration, light intensity, and swimming orientation are strongly coupled through nonlinear feedback mechanisms. 

In the former case, nonlinear effects saturate the disturbance growth, leading to stable finite-amplitude convection \citep{watson1960non,stuart1960non,malkus1958finite}, whereas subcritical bifurcations may support finite-amplitude states below the linear threshold, giving rise to hysteresis and bistability. A standard approach for characterizing such behaviour is the weakly nonlinear method developed by \citet{watson1960non}, \citet{stuart1960non}, and \citet{malkus1958finite}. In this framework, the control parameter is expanded about its critical value, and a slow temporal scale is introduced to describe the gradual evolution of the disturbance amplitude. At third order, the analysis yields the Stuart–Landau amplitude equation \citep{landau1944problem}, whose coefficients determine the nature of the bifurcation. Owing to its systematic extension of linear stability theory, this methodology has been widely employed in the study of convection-driven instabilities \citep{ragoju2020linear,lakshmi2022linear,an1995weakly,bouteraa2015weakly,siddheshwar2025linear,bhadauria2012weakly,bhadauria2013effect,kiran2016weakly,skene2024weakly,castelino2018linear}.

Building upon these developments, the present work extends the weakly nonlinear framework to thermally driven phototactic bioconvection under collimated irradiation in a fluid layer heated or cooled from below. The phototaxis model of \citet{kumar2024thermal} is adopted, in which the incompressible Navier–Stokes equations are coupled with the temperature equation and a conservation equation for the motile microorganism population. Particular emphasis is placed on the influence of thermal buoyancy on the post-critical evolution of disturbances and the associated bifurcation behaviour. Such investigations are of practical importance in bioremediation, wastewater treatment, and biomass cultivation, where bioconvection enhances microorganism transport, improves nutrient and light distribution, and reduces the need for mechanical mixing \citep{barsanti2025bioconvection}. Furthermore, while directed illumination governs phototactic migration, thermal buoyancy provides an additional driving mechanism that significantly modifies convective flow and pattern formation \citep{thiffeault2010stirring}.

 The paper is structured as follows. The governing equations and mathematical formulation are introduced in Section \ref{sec2}, followed by the derivation of the basic equilibrium state in Section \ref{sec3}. Sections \ref{sec4}-\ref{sec6} are devoted to the perturbation analysis and the development of the weakly nonlinear theory. The resulting numerical findings, together with their physical significance, are presented in Section \ref{sec7}. Finally, the conclusion drawn from the study are summarized in Section \ref{sec8}.

\begin{figure*}[b]
 	\begin{center}
 			\includegraphics[ width=0.9\textwidth, height=0.35\textheight,trim=6cm 18cm 3cm 5cm,
    clip]{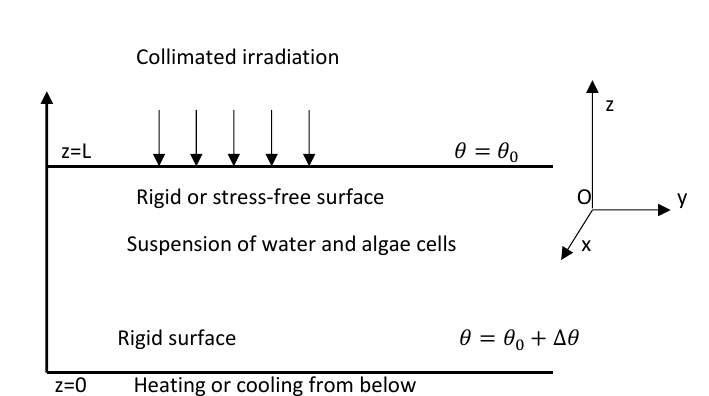}	
             \caption{Schematic representation of the thermally influenced phototactic bioconvection system.}
   \label{Collimated irradiation.pdf}
 		\end{center}       
 \end{figure*}
\section{Governing Equations}
\label{sec2}

Consider a horizontal layer of an incompressible viscous microbial suspension of depth $L$, confined between two parallel boundaries located at $z=0$ and $z=L$ . A Cartesian coordinate system is employed with the $z$-axis directed vertically upward from the lower boundary. The lower boundary is assumed to be rigid, while the upper boundary may be either rigid or stress-free, depending on the boundary conditions under consideration. The suspension is illuminated from above by a collimated light source, whereas the lower boundary is maintained at a temperature of $\theta_0+\Delta \theta$. The upper boundary is kept at a constant temperature $\theta_0$. As a result, a temperature difference $\Delta \theta$ is established across the layer, corresponding to either heating from below $(\Delta \theta>0)$ or cooling from below $(\Delta \theta<0)$.

Under the Boussinesq approximation, the evolution of the suspension is described by the conservation of mass, momentum, thermal energy, and microorganism concentration. The corresponding governing equations are given by
\begin{align}
    &\boldsymbol{\nabla}\cdot\boldsymbol{u}=0,\\
    &\left(\frac{\partial}{\partial t}
    +\boldsymbol{u}\cdot\boldsymbol{\nabla}\right)\boldsymbol{u}
    =\frac{1}{\rho}
    \left[-\boldsymbol{\nabla}\mathbb{P}
    -n\vartheta\Delta\rho\,g\,\hat{k}
    +\mu\nabla^{2}\boldsymbol{u}
    +\daleth(\theta-\theta_{0})g\,\hat{k}\right],\\
    &\rho c
    \left(\frac{\partial}{\partial t}
    +\boldsymbol{u}\cdot\boldsymbol{\nabla}\right)\theta
    =k\nabla^{2}\theta,\\
    &\frac{\partial n}{\partial t}
    +\boldsymbol{\nabla}\cdot
    \left[n(\boldsymbol{u}+\boldsymbol{U})
    -\boldsymbol{D}\boldsymbol{\nabla}n\right]=0.
\end{align}

The quantity $\rho c$ denotes the volumetric heat capacity of the fluid, where $\rho$ is the fluid density and $c$ is the specific heat capacity. The thermal conductivity is represented by $k$, and the corresponding thermal diffusivity is defined as $\beta_f=k/(\rho c)$. The variable $n$ represents the cell concentration of swimming microorganisms. Each microorganism occupies a volume $\vartheta$ and possesses a density $\rho+\Delta\rho$, with the relative density difference satisfying $\Delta\rho/\rho \ll 1$. The quantity, $\boldsymbol{u}$ is the fluid velocity field, while $\mu$ denotes the dynamic viscosity. The acceleration due to gravity acts in the negative vertical direction and is denoted by $g$, whereas $\mathbb{P}$ corresponds to the pressure field. The transport of microorganisms arises from advection by the fluid motion, active swimming, and diffusive spreading. Accordingly, the total cell flux can be written as $\boldsymbol{M} =n(\boldsymbol{u}+\boldsymbol{U}) -\boldsymbol{D}\boldsymbol{\nabla}n$. For simplicity, the diffusivity tensor is assumed to be homogeneous and isotropic throughout the suspension, such that $\boldsymbol{D}=DI$, where $D$ is the cell diffusivity and $I$ denotes the identity tensor. The swimming velocity of the microorganisms is given by
\[
\boldsymbol{U}=U\hat{p},
\]
where $U$ is the mean swimming speed and $\hat{p}$ denotes the average cell swimming orientation vector. Following \citet{ref14}, the orientation vector is given as
\begin{eqnarray}
\label{eq7}
\hat{p}=T(K)\frac{\boldsymbol{G}}{|\boldsymbol{G}|}.
\end{eqnarray}

Here $T(K)$ characterizes the phototactic response function of the cells and depends on the total light intensity $K(\boldsymbol{x})
=\int_{0}^{4\pi}
J(\boldsymbol{x},\boldsymbol{q})\,d\Omega$. The radiative flux vector at a spatial location $\boldsymbol{x}=(x,y,z)$ is obtained from the angular distribution of the radiative intensity as
\begin{equation}
\boldsymbol{G}
=\int_{0}^{4\pi}
J(\boldsymbol{x},\boldsymbol{q})\,\boldsymbol{q}\,d\Omega=|\boldsymbol{G}|\,\hat{k},
\end{equation}
and Eq.~(\ref{eq7}) reduces to
\begin{equation}
\label{eq9}
\hat{p}=T(K)\hat{k}.
\end{equation}
Here $J(\boldsymbol{x},\boldsymbol{q})$ denotes the radiative intensity in the direction specified by the unit vector $\boldsymbol{q}
=\sin\gamma_1
\cos\gamma_2\,\hat{i}
+\sin\gamma_1 sin\gamma_2\,\hat{j}
+\cos\gamma_1\,\hat{k}$, where $\gamma_1$ and $\gamma_2$ denote the polar and azimuthal angles, respectively, and $\hat{i}$, $\hat{j}$, and $\hat{k}$ are the unit vectors along the $x$, $y$, and $z$-directions. To determine the radiative intensity within the suspension, the transport of light is described by the radiative transfer equation (RTE) for a purely absorbing, non-scattering medium \cite{ref-modest,ref17},
\begin{equation}
\boldsymbol{q}\cdot\nabla J
=-\xi J.
\end{equation}
The absorption coefficient $\xi$ is assumed to depend on the algae cells concentration through the linear relation $\xi=\varpi n$, where $\varpi$ is a constant absorption parameter.  The upper and lower surfaces are taken to be non-reflecting. Accordingly, the radiative boundary conditions are given by
\begin{align}
&J(x,y,0,\gamma_1,\gamma_2)=
0,\quad   \gamma_1\in[0,\pi/{2}],\\[2mm]
&J(x,y,L,\gamma_1,\gamma_2)=J_s\,\delta(\boldsymbol{q}-\boldsymbol{q}_s),
\quad  \gamma_1\in[\pi/{2},\pi],
\end{align}
where $J_s$ denotes the light intensity of the incident radiation, $\boldsymbol{q}_s=-\hat{k}$ specifies the downward direction of the incoming beam, and $\delta$ represents the Dirac delta function. The solution of the radiative transfer problem provides the radiative intensity $J$, from which the total light intensity experienced by the microorganisms can be evaluated as
\begin{equation}
K(\boldsymbol{x})
=\int_{0}^{4\pi}
J(\boldsymbol{x},\boldsymbol{q})\,d\Omega.
\end{equation}
Since the swimming behaviour of phototactic microorganisms is regulated by the local illumination level, the function $T(K)$ appearing in Eq.~(\ref{eq9}) is expressed as \cite{ref13,ref17}
\begin{align}
T(K)
=&\,0.8\sin\!\left(1.5\Lambda\right)
-0.1\sin\!\left(0.5\Lambda\right),\nonumber\\
\Lambda
=&\,\pi K e^{\Upsilon(K-1)}.
\end{align}
Here, the parameter $\Upsilon$ is related to the critical light intensity $K_c$ and controls the sensitivity of the phototactic response to variations in illumination.


To close the governing system, suitable boundary conditions are imposed on the velocity field and microorganism flux at the horizontal boundaries $z=0$ (bottom) and $z=L$ (top). Since both boundaries are impermeable, the normal velocity component vanishes. Thus,
\[
\boldsymbol{u}\cdot\hat{k}=w=0,
\qquad \text{at} \quad z=0, \quad \text{and} \quad L.
\]
It is assumed that the suspension is heated/cooled from the bottom via constant temperature. The temperature at the bottom of the suspension is set to $\theta_0 + \Delta \theta$, whereas $\theta_0$ represents the temperature at the top boundary.  Thus, the thermal boundary conditions are
\[
\theta=\theta_0 + \Delta \theta \qquad \text{at} \quad z=0,\qquad 
\theta=\theta_0 \qquad \text{at} \quad z=1.
\]
The absence of microorganism transport through the boundaries further requires
\[
\boldsymbol{M}\cdot\hat{k}=0,
\qquad \text{at} \quad z=0, \quad \text{and} \quad L.
\]
The boundaries are modeled using either rigid or stress-free conditions, depending on the mechanical characteristics of the surfaces. For a stress-free surface,
\[
\frac{\partial^2}{\partial z^2}(\boldsymbol{u}\cdot\hat{k})=0,
\]
whereas for a rigid surface,
\[
\boldsymbol{u}\times\hat{\boldsymbol{z}}=0.
\]


To express the governing equations in dimensionless form, the characteristic scales $\frac{\mu \beta_f}{L^{2}}$, $L$, $\frac{\beta_f}{L}$, $\frac{L^{2}}{\beta_f}$, $\Delta \theta$, $\bar n$ are selected for pressure, length, velocity, time, temperature difference, and microorganism concentration, respectively. Employing these scales, the governing equations become
\begin{align}
&\boldsymbol{\nabla}\cdot\boldsymbol{u}=0,\\
&\frac{1}{P_r}
\left(
\frac{\partial \boldsymbol{u}}{\partial t}
+\boldsymbol{u}\cdot\boldsymbol{\nabla}\boldsymbol{u}
\right)
=
-\boldsymbol{\nabla}\mathbb{P}
-nR_n\hat{k}
+\nabla^{2}\boldsymbol{u}
+ R_t\theta\hat{k},\\
&\left(
\frac{\partial}{\partial t}
+\boldsymbol{u}\cdot\boldsymbol{\nabla}
\right)\theta
=
\nabla^{2}\theta,\\
&\frac{\partial n}{\partial t}
=
\boldsymbol{\nabla}\cdot
\left[
\frac{1}{L_n}\boldsymbol{\nabla}n
-n\Bigl(
\boldsymbol{u}
+\frac{1}{L_n} VT(K)\hat{k}
\Bigr)
\right].
\end{align}

The behaviour of the system is governed by several dimensionless parameters, defined as
\[
\nu=\frac{\mu}{\rho},
\qquad
R_n=\frac{\bar n\vartheta\Delta\rho gL^{3}}
{\mu \beta_f},
\quad
L_n=\frac{\beta_f}{D},
\qquad
V=\frac{UL}{D},
\qquad
P_r=\frac{\mu}{\rho \beta_f},
\qquad
R_t=
\frac{\daleth\,\Delta T\,\rho gL^{3}}
{\mu\beta_f}.
\]

Here, $R_n$ measures the bioconvection Rayleigh number. The parameter $R_t$ represents the thermal Rayleigh number, quantifying the strength of thermally induced buoyancy. The Lewis number $L_n$ compares thermal diffusion with microorganism diffusion, while $V$ characterizes the cell swimming speed relative to diffusive transport. $P_r$ is the Prandtl number and $\nu$ is the kinematic viscosity of the fluid. The dimensionless boundary conditions are given by
\begin{align*}
\boldsymbol{u}\cdot\hat{k}
=
0,
\qquad
\left[
n\Bigl(\frac{1}{L_n}
VT(K)\hat{k}
+\boldsymbol{u}
\Bigr)
-\frac{1}{L_n}\boldsymbol{\nabla}n
\right]\cdot\hat{k}
=
0,
\qquad z=0,1,
\end{align*}
\begin{align*}
\theta=1 \quad \text{at} \quad z=0,
\qquad
\theta=0 \quad \text{at} \quad z=1.
\end{align*}

For a rigid boundary,
\begin{align*}
\boldsymbol{u}\times\hat{k}=0,
\qquad z=0,1.
\end{align*}
Alternatively, when the upper surface is stress-free,
\begin{align*}
\frac{\partial^{2}}
{\partial z^{2}}
\left(
\boldsymbol{u}\cdot\hat{k}
\right)
=0,
\qquad z=1.
\end{align*}

The nondimensional radiative transfer equation is
\begin{equation}
\label{eqn:equation10}
\frac{dJ}{dq}
+\kappa Jn
=0.
\end{equation}
The corresponding boundary conditions are
\begin{align}
\label{eqn:equation11}
&J(x,y,0,\gamma_1,\gamma_2)
=
0,
\qquad
\gamma_1\in[0,\pi/2],\\[2mm]
&J(x,y,1,\gamma_1,\gamma_2)
=
J_s\,\delta(\boldsymbol{q}-\boldsymbol{q}_s),
\qquad
\gamma_1\in[\pi/2,\pi].
\end{align}

The parameter $\kappa=\varpi\bar nL$ represents the optical thickness of the suspension layer and provides a measure of the cumulative attenuation experienced by light as it traverses the medium. Larger values of $\kappa$ correspond to stronger absorption and consequently a more pronounced influence of the microorganism distribution on the internal light field.

\section{Steady state solution}
\label{sec3}
The basic-state temperature distribution is given by
\[
\theta_0 = 1-z.
\]

For the quiescent basic state, the cell conservation equation for the microorganisms reduces to
\begin{equation}
\label{eqn:equation12}
\frac{dn_0}{dz}=VT(K_0)\,n_0,
\end{equation}
subject to the normalization condition
\begin{equation}
\label{eqn:equation13}
\int_0^1 n_0(z)\,dz = 1.
\end{equation}

Here, $n_0$ denotes the steady-state cell concentration, and $K_0 = J_s \exp\!\left(\kappa\int_1^z n_0(z')\,dz'\right)$ is the corresponding light intensity. The solution of the radiative transfer equation is
\[
J_0
=
J_s \exp\!\left(-\kappa\int_1^z n_0(z')\,dz'\right)
\delta(\boldsymbol{q}-\boldsymbol{q}_s).
\]

Introducing the new variable $\zeta(z)=\int_1^z n_0(z')\,dz'$, so that $n_0=\frac{d\zeta}{dz}$. Now, Eq.~\eqref{eqn:equation12} can be rewritten as
\begin{equation}
\label{basicstate}
\frac{d^2\zeta}{dz^2}
-
VT(K_0)\frac{d\zeta}{dz}
=0,
\end{equation}
where
\[
K_0 = J_s e^{\kappa\zeta}.
\]

The corresponding boundary conditions are
\begin{equation}
\label{basicstatebc}
\zeta(0)=-1,
\qquad
\zeta(1)=0.
\end{equation}

The resulting boundary-value problem is solved numerically to determine the steady-state cell concentration $n_0(z)$.

\section{First order problem}
\label{sec4}

To obtain the amplitude equation, the dependent variables are expressed as asymptotic expansions in terms of a small parameter $\chi$. Accordingly, the velocity, temperature, and cell concentration fields are expanded as $\boldsymbol{u}=\chi \boldsymbol{u}_1+\chi^2 \boldsymbol{u}_2+\chi^3 \boldsymbol{u}_3+\cdots$, $\theta=\theta_0+\chi \theta_1+\chi^2 \theta_2+\chi^3 \theta_3+\cdots$, $n=n_0+\chi n_1+\chi^2 n_2+\chi^3 n_3+\cdots$, with analogous expansions assumed for $\mathbb{P}$, $K$, and $\hat{p}$. Since the analysis is performed in the vicinity of the critical state, the control parameters are also expanded in powers of $\chi$. Thus, the bioconvection and thermal Rayleigh numbers are written as $R_n=R_0+\chi^2R_2+\chi^4R_4+\cdots$, and $R_t=R_{t0}+\chi^2 R_{t2}+\chi^4 R_{t4}+\cdots$, respectively, where $R_0$ denotes the critical bioconvection Rayleigh number corresponding to the onset of instability, while $R_{t0}$ denotes the thermal Rayleigh number. The expansion of both $R_n$ and $R_t$ enables the coupled effects of bioconvective and thermal buoyancy to be incorporated within a unified weakly nonlinear framework. In particular, the formulation reduces to the pure bioconvective problem when $R_{t0}=0$. As $R_{t0}$ is increased and approaches the critical thermal Rayleigh number associated with the onset of thermal convection, the formulation consistently recovers the classical Rayleigh--B\'enard convection problem. To account for the slow modulation of the disturbance amplitude near criticality, a multiple-time-scale formulation is introduced. The temporal derivative is therefore decomposed as $\frac{\partial}{\partial t}=\frac{\partial}{\partial \tau_f}+\chi^2\frac{\partial}{\partial \tau_s}$, where $\tau_f$ and $\tau_s$ represent the fast and slow temporal scales, respectively. This multiple-time-scale approach separates the fast evolution of the perturbation fields from the slow modulation of the amplitude near the onset of instability.

At the leading order, \(O(\chi)\), the nonlinear contributions are absent and the governing equations reduce to a linear eigenvalue problem describing the onset of convection. The resulting system is
\begin{align}
\label{eqn:equation15}
&\nabla \cdot \boldsymbol{u}_1 =0,\\
\label{eqn:equation16}
&\frac{\partial \boldsymbol{u}_1}{\partial \tau_f}
-P_r(
-\boldsymbol{\nabla} \mathbb{P}_1
-n_1 R_0 \hat{k}
+\nabla^2\boldsymbol{u}_1
+R_{t0}\theta_1\hat{k}
)=0,\\
&\frac{\partial \theta_1}{\partial \tau_f}
-w_1
-\nabla^2\theta_1=0,\\
\label{eqn:equation17}
&\frac{\partial n_1}{\partial \tau_f}
+\frac{dn_0}{dz}w_1
+\frac{1}{L_n} V\nabla\cdot\left(n_1\hat{p}_0+n_0\hat{p}_1\right)
-\frac{1}{L_n}\nabla^2 n_1=0.
\end{align}

To remove the pressure term from the momentum equation, the curl operator is applied twice to Equation ~(\ref{eqn:equation16}), and the vertical component of the resulting expression is retained. This yields
\begin{equation}
\label{eqn:equation20}
\nabla^4 w_1
-\frac{1}{P_r}
\frac{\partial}{\partial \tau_f}
\left(\nabla^2 w_1\right)
-R_0(\partial_{xx}+\partial_{yy})n_1
+R_{t0}(\partial_{xx}+\partial_{yy})\theta_1
=0,
\end{equation}
where
\[
\partial_{xx}=\frac{\partial^2}{\partial x^2},
\qquad
\partial_{yy}=\frac{\partial^2}{\partial y^2}.
\]

The perturbation fields satisfy the boundary conditions
\begin{align*}
&w_1=\theta_1=0,
\qquad z=0,1,\quad \frac{\partial n_1}{\partial z}
-VT_0n_1=0,
\qquad z=1.\\
&\frac{\partial n_1}{\partial z}
-VT_0n_1
+\kappa V n_0K_0\frac{dT_0}{dK_0}
\int_z^1 n_1(x,y,z',t)\,dz'
=0,
\qquad z=0.
\end{align*}

For rigid boundaries, the no-slip condition further requires
\[
\frac{\partial w_1}{\partial z}=0,
\qquad z=0,1,
\]
whereas for a stress-free boundary,
\[
\frac{\partial^2 w_1}{\partial z^2}=0,
\qquad z=1.
\]

To facilitate the analysis, the transformed variable $n_1'(x,y,z,t)=\int_z^1 n_1(x,y,z',t)\,dz'$ is introduced. In terms of \(n_1'\), the linearized system of equations becomes
\begin{align}
\label{eqn22}
&\nabla^4w_1
-\frac{1}{P_r}
\frac{\partial}{\partial\tau_f}
(\nabla^2w_1)
+R_0(\partial_{xx}+\partial_{yy})
\frac{\partial n_1'}{\partial z}
+R_{t0}
(\partial_{xx}+\partial_{yy})\theta_1
=0,
\\
\label{eqn23}
&
\frac{\partial^2 n_1'}
{\partial\tau_f\partial z}
-\frac{1}{L_n}\nabla^2
\frac{\partial n_1'}{\partial z}
+\frac{V}{L_n} \left[
T_0\frac{\partial^2}{\partial z^2}
+2\kappa n_0K_0
\frac{\partial T_0}{\partial K_0}
\frac{\partial}{\partial z}
+\kappa\frac{\partial}{\partial z}
\left(
n_0K_0\frac{dT_0}{dK_0}
\right)
\right]n_1'
-\frac{\partial n_0}{\partial z}w_1
=0,
\\
&
\nabla^2\theta_1
-\frac{\partial\theta_1}{\partial\tau_f}
+w_1
=0.
\end{align}

The above equations are written in the compact operator form as
\begin{equation}
\label{eqn:equation21}
\mathcal{L}\boldsymbol{X}_1=\boldsymbol{f}_1,
\end{equation}
or equivalently,
\begin{equation}
\label{eqn25}
\begin{bmatrix}
\mathcal{L}_{11}
&
\mathcal{L}_{12}
&
\mathcal{L}_{13}
\\
-L_n\dfrac{\partial n_0}{\partial z}
&
L_n\dfrac{\partial^2}{\partial\tau_f\partial z}-\digamma
&
0
\\
1
&
0
&
\nabla^2-\dfrac{\partial}{\partial\tau_f}
\end{bmatrix}
\begin{bmatrix}
w_1\\
n_1'\\
\theta_1
\end{bmatrix}
=
\begin{bmatrix}
f_{11}\\
f_{12}\\
f_{13}
\end{bmatrix}.
\end{equation}

Since the first-order problem is homogeneous, $\boldsymbol{f}_1
=
\begin{bmatrix}
f_{11}&f_{12}&f_{13}
\end{bmatrix}^{T}
=
\begin{bmatrix}
0&0&0
\end{bmatrix}^{T}$. The operators appearing in Equation~(\ref{eqn25}) are defined by
\begin{align*}
&\mathcal{L}_{11}
=
\nabla^4
-\frac{1}{P_r}
\frac{\partial}{\partial\tau_f}
\nabla^2,
\quad
\mathcal{L}_{12}
=
R_0
(\partial_{xx}+\partial_{yy})
\frac{\partial}{\partial z},
\quad
\mathcal{L}_{13}
=
R_{t0}
(\partial_{xx}+\partial_{yy}),
\\
&\digamma
=
\nabla^2\frac{\partial}{\partial z}
-
V\left[
T_0\frac{\partial^2}{\partial z^2}
+
2\kappa n_0K_0
\frac{\partial T_0}{\partial K_0}
\frac{\partial}{\partial z}
+
\kappa\frac{\partial}{\partial z}
\left(
n_0K_0\frac{dT_0}{dK_0}
\right)
\right].
\end{align*}

To determine the linear stability characteristics of the system, solutions are sought in the form of normal modes. Accordingly, the perturbation fields are represented by
\begin{align}
\label{eqn:equation22}
w_1&=
A(\tau_s)\,W_1(z)
e^{i(\lambda_1 x+\lambda_2 y)+\eta\tau_f}
+c.c.,\\
\label{eqn:equation23}
n_1'&=
A(\tau_s)\,N_1'(z)
e^{i(\lambda_1x+\lambda_2y)+\eta\tau_f}
+c.c.,\\
\label{eqn:equation231}
\theta_1&=
A(\tau_s)\,\Theta_1(z)
e^{i(\lambda_1x+\lambda_2y)+\eta\tau_f}
+c.c.,
\end{align}
where \(A(\tau_s)\) denotes the slowly varying amplitude, \(\lambda=\sqrt{\lambda_1^2+\lambda_2^2}\) is the horizontal wavenumber, \(\eta=\eta_r+i\eta_i\) represents the complex growth rate, and \(c.c.\) denotes the complex conjugate of the preceding term. The perturbation in the microorganism concentration can likewise be expressed as $n_1=
A(\tau_s)\,N_1(z)
e^{i(\lambda_1x+\lambda_2y)+\eta\tau_f}
+c.c.$. Substitution of the normal-mode expansions into the linearized governing equations yields the following system of ordinary differential equations:
\begin{align}
\label{eqn:equation24}
&
\begin{bmatrix}
\left(
\dfrac{d^2}{dz^2}-m^2
\right)^2
&
-m^2 R_0\dfrac{d}{dz}
&
-m^2 R_{t0}
\\
L_n \dfrac{dn_0}{dz}
&
\digamma_1
&
0
\\
1
&
0
&
\left(
\dfrac{d^2}{dz^2}-m^2
\right)
\end{bmatrix}
\begin{bmatrix}
W_1\\
N_1'\\
\Theta_1
\end{bmatrix}
=
\eta
\begin{bmatrix}
\dfrac{1}{P_r}
\left(
\dfrac{d^2}{dz^2}-m^2
\right)
&
0
&
0
\\
0
&
L_n\dfrac{d}{dz}
&
0
\\
0
&
0
&
1
\end{bmatrix}
\begin{bmatrix}
W_1\\
N_1'\\
\Theta_1
\end{bmatrix}.
\end{align}

The differential operator \(\digamma_1\) appearing above is defined by
\begin{align*}
\digamma_1
=
\frac{d^3}{dz^3}
-VT_0\frac{d^2}{dz^2}
-
\left(
m^2
+
2\kappa V n_0K_0
\frac{dT_0}{dK_0}
\right)
\frac{d}{dz}
-\kappa V
\frac{d}{dz}
\left(
n_0K_0
\frac{dT_0}{dK_0}
\right).
\end{align*}

The corresponding boundary conditions become
\begin{align}
\label{eqn:equation25}
&W_1=\Theta_1=0,
\qquad z=0,1,
\nonumber\\
&
\frac{d^2N_1'}{dz^2}
-\kappa V n_0K_0
\frac{dT_0}{dK_0}
N_1'
-VT_0
\frac{dN_1'}{dz}
=0,
\qquad z=0,
\\
&
\frac{d^2N_1'}{dz^2}
-VT_0
\frac{dN_1'}{dz}
=0,
\quad
N_1'=0,
\qquad z=1.
\end{align}

For rigid boundaries,
\begin{equation}
\frac{dW_1}{dz}=0,
\qquad z=0,1.
\end{equation}

In contrast, when the upper boundary is stress-free,
\begin{equation}
\label{eqn:equation28}
\frac{d^2W_1}{dz^2}=0,
\qquad z=1.
\end{equation}

The differential system given by Equation~(\ref{eqn:equation24}), together with the associated boundary conditions, constitutes a generalized eigenvalue problem of the form $A_g\boldsymbol{X}_{1g} =\eta B_g\boldsymbol{X}_{1g}$, where the eigenvector is $\boldsymbol{X}_{1g}=\begin{bmatrix}W_1 &N_1'&\Theta_1\end{bmatrix}^T$.



 \begin{figure*}[b]
 	\begin{center}
 			\includegraphics[ width=0.9\textwidth, height=0.35\textheight]{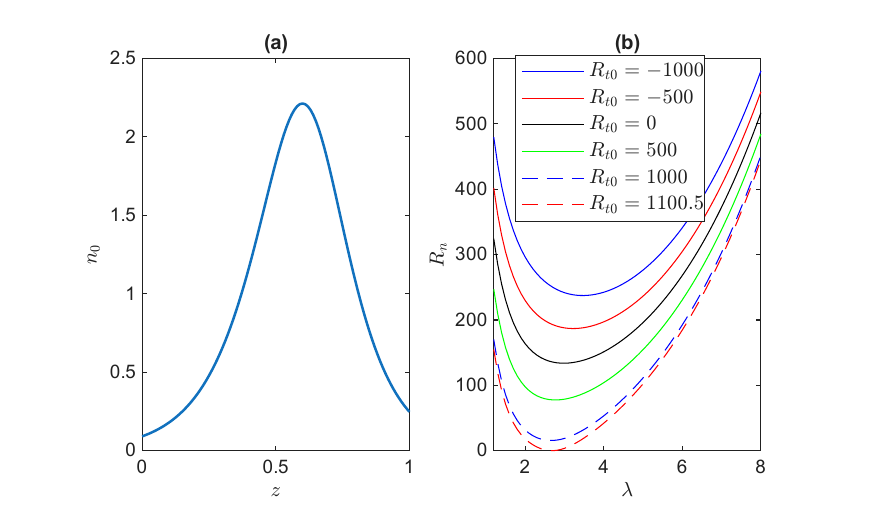}	
             \caption{(a) Basic-state microorganism concentration profile for $V=10$, $K_c=0.5$, and $\kappa=1$. (b) Neutral stability curves for various values of the thermal Rayleigh number $R_{t0}$ under rigid-free boundary conditions.}
   \label{10v1.00.5Iclinear+basic.pdf}
 		\end{center}       
 \end{figure*}

\begin{figure*}[b]
	\begin{center}
			\includegraphics[ width=1\textwidth, height=0.35\textheight]{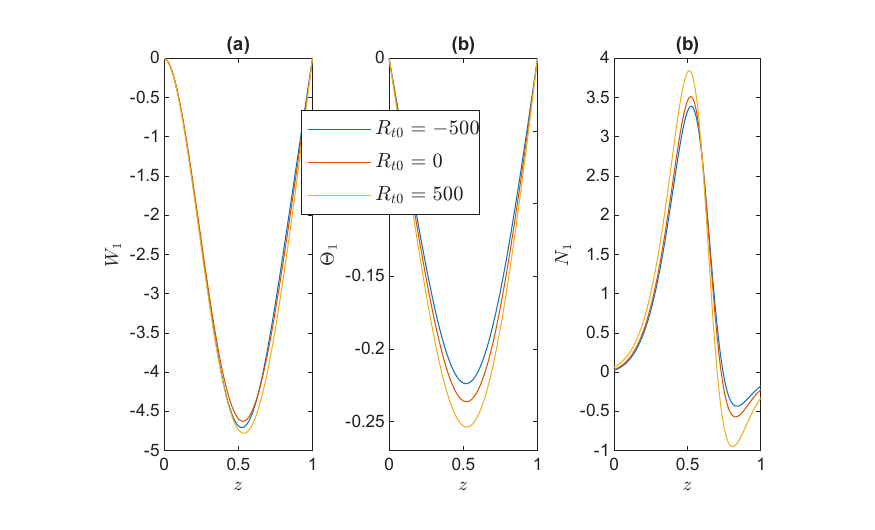}	
            \caption{Eigenfunctions of the linear stability analysis at critical states: (a) vertical velocity component, (b) vorticity field, and (c) cell concentration distribution across various Taylor numbers $T_a$. The remaining parameters are held constant at $V = 10$, $K_c = 0.5$, and $\kappa = 1$.}
   \label{fig:linearsol.pdf}
		\end{center}       
\end{figure*}
\begin{figure*}[b]
	\begin{center}
			\includegraphics[ width=1\textwidth, height=0.35\textheight]{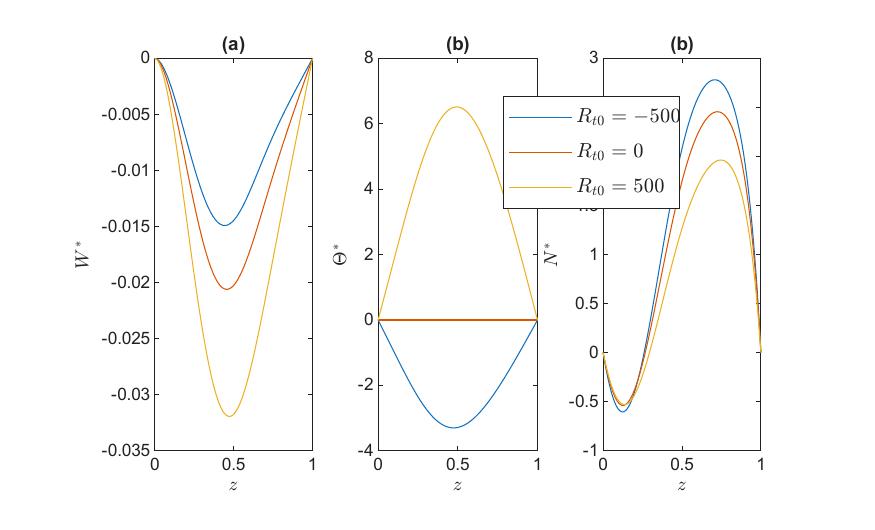}	
            \caption{Adjoint eigenfunctions evaluated at critical points: (a) vertical velocity of the fluid, (b) vorticity, and (c) distribution of cell concentration for various Taylor numbers $T_a$. The analysis is conducted with fixed parameter values of $V = 10$, $K_c = 0.5$, and $\kappa = 1$.}
   \label{fig:adjsol.pdf}
		\end{center}       
\end{figure*}
\begin{figure*}[b]
	\begin{center}
			\includegraphics[ width=1\textwidth, height=0.35\textheight]{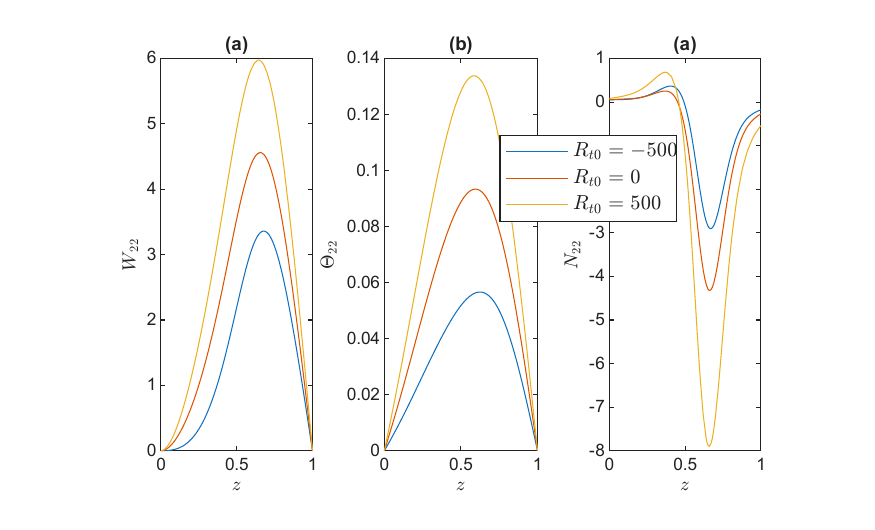}	
            \caption{At critical conditions, first-harmonic components are associated with the fundamental mode: (a) vertical fluid velocity, (b) vorticity, and (c) cell concentration for different Taylor numbers $T_a$, with $V = 10$, $K_c = 0.5$, and $\kappa = 1$.}
   \label{fig:secorder2sol.pdf}
		\end{center}       
\end{figure*}
\begin{figure*}[b]
	\begin{center}
			\includegraphics[ width=1\textwidth, height=0.35\textheight]{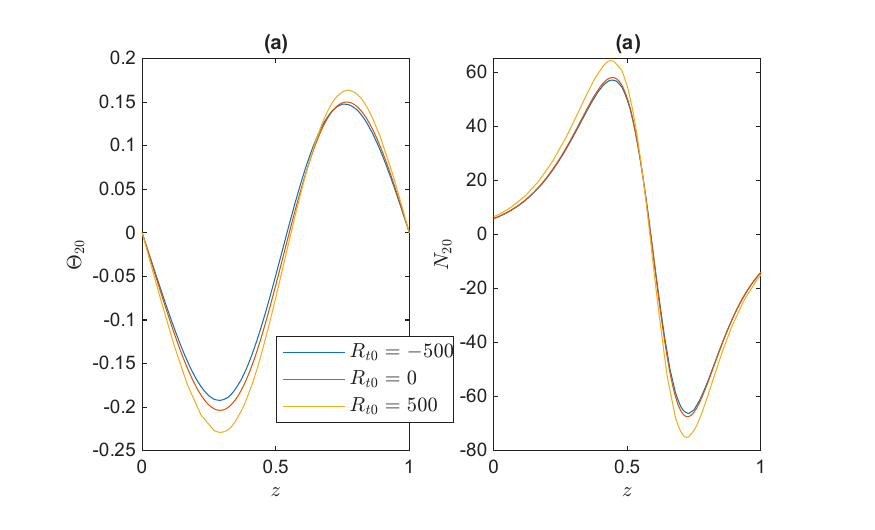}	
            \caption{At critical conditions, corrections to the basic-state cell concentration profiles: (a) vertical fluid velocity, (b) vorticity, and (c) cell concentration for different Taylor numbers $T_a$, with $V = 10$, $K_c = 0.5$, and $\kappa = 1$.}
   \label{fig:secorder20sol.pdf}
		\end{center}       
\end{figure*}
\begin{figure*}[t]
    \centering
    
    \begin{subfigure}{0.49\textwidth}
        \centering
        \includegraphics[trim=1cm 0.5cm 1cm 0.6cm,
    clip, width=\linewidth, height=0.35\textheight, keepaspectratio]{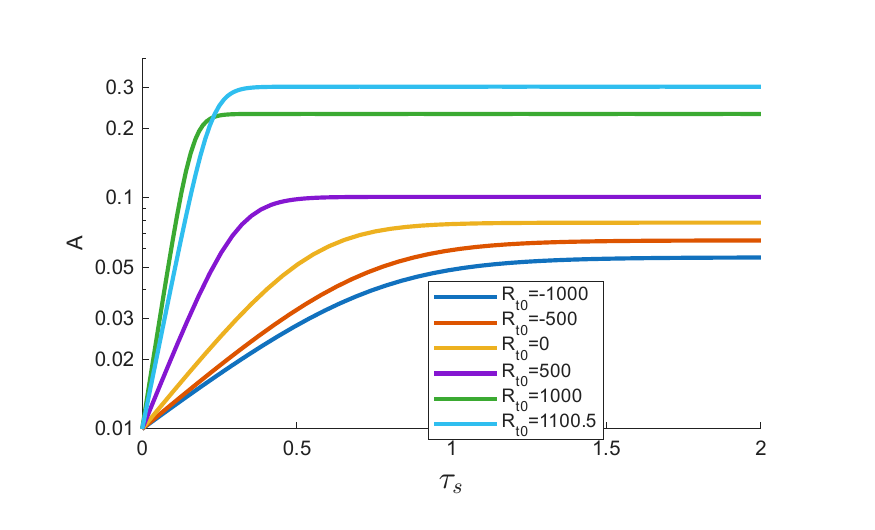}
        \caption{}
        \label{fig:Amplitude10v1k0.5Ic}
    \end{subfigure}
    \hfill
    \begin{subfigure}{0.49\textwidth}
        \centering
        \includegraphics[trim=0.5cm 0.2cm 1cm 0.3cm,
    clip, width=\linewidth, height=0.35\textheight, keepaspectratio]{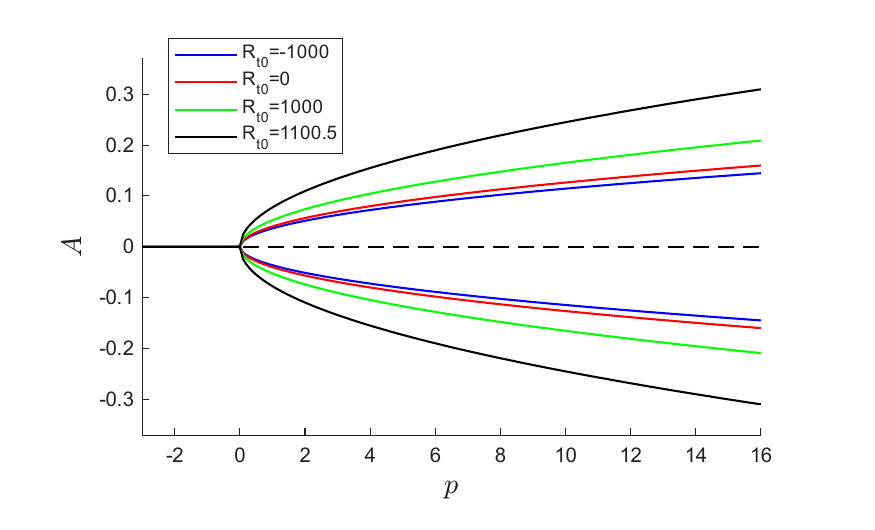}
        \caption{}
        \label{fig:bifurcation10v1k0.5Icfree.pdf}
    \end{subfigure}
    
    \caption{(a) Temporal evolution of the amplitude $|A|$ with slow time $s$ for different Taylor numbers $T_a$, with fixed parameters $V = 15$, $\kappa = 1$, and $K_c = 0.51$. 
(b) Corresponding supercritical Hopf bifurcation diagrams for Taylor numbers $T_a=800$.}
    \label{fig:amp+pitchfork}
\end{figure*}

\begin{figure*}[b]
	\begin{center}
			\includegraphics[ width=1\textwidth, height=0.35\textheight]{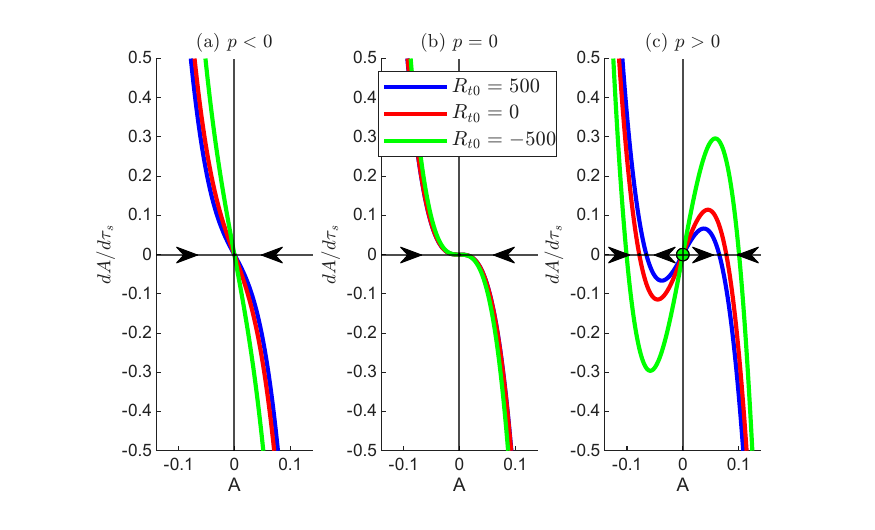}	
            \caption{Phase-plane vector fields for different values of the parameter $a$ under varying Taylor numbers $T_a$, with $K_c = 0.64$, $V = 15$, and $\kappa = 0.5$ fixed. The blue curves represent $T_a = 0$, while the red curves correspond to $T_a = 1000$.}
   \label{fig:vectorfield10v1k0.5Ic.pdf}
		\end{center}       
\end{figure*}
\begin{figure*}[b]
	\begin{center}
			\includegraphics[ width=1\textwidth, height=0.35\textheight]{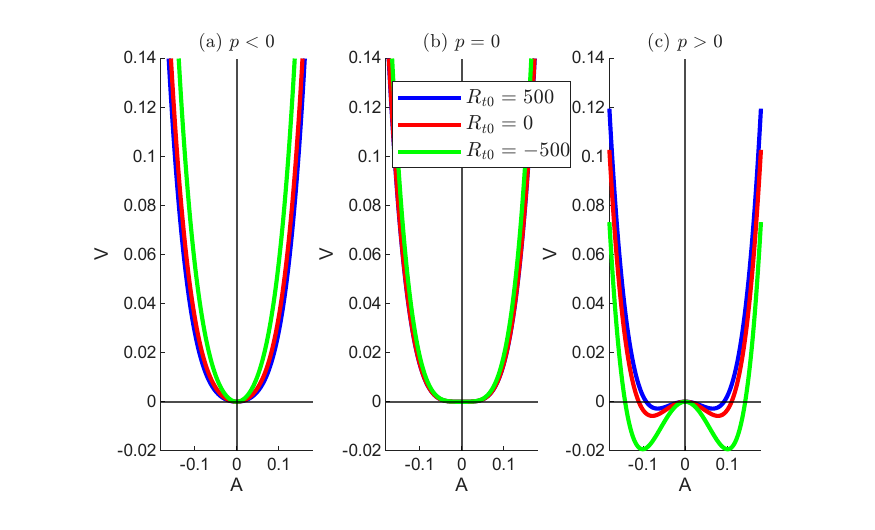}	
            \caption{Potential function $V(A)$  for different values of the parameter $a$ under varying Taylor numbers $T_a$, with $K_c = 0.64$, $V = 15$, and $\kappa = 0.5$ fixed. The blue curves represent $T_a = 0$, while the red curves correspond to $T_a = 1000$.}
   \label{fig:potential10v1k0.5Ic.pdf}
		\end{center}       
\end{figure*}

\begin{figure*}[b]
	\begin{center}
			\includegraphics[ width=1\textwidth, height=0.35\textheight]{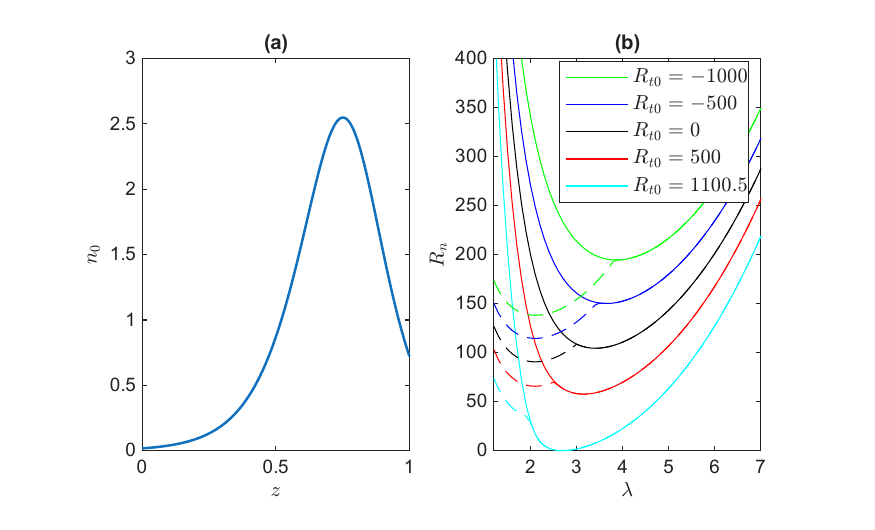}	
            \caption{(a) Basic-state microorganism concentration profile for $V=15$, $K_c=0.65$, and $\kappa=0.5$. (b) Neutral stability curves for various values of the thermal Rayleigh number $R_{t0}$ under rigid-free boundary conditions.}
   \label{fig:15v0.5k0.65IcRF.pdf}
		\end{center}       
\end{figure*}

\begin{figure*}[t]
    \centering
    
    \begin{subfigure}{0.49\textwidth}
        \centering
        \includegraphics[trim=1cm 0.5cm 1cm 0.6cm,
    clip, width=\linewidth, height=0.35\textheight, keepaspectratio]{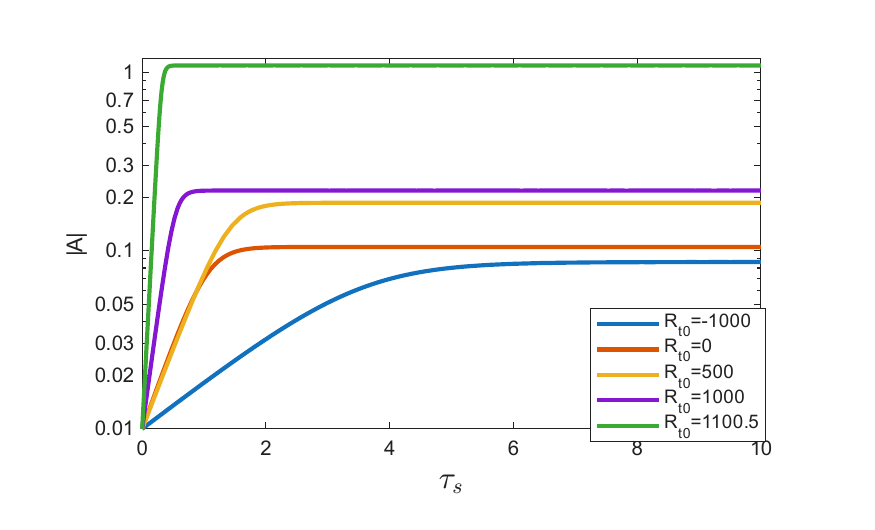}
        \caption{}
        \label{fig:Amplitude15v0.5k0.65Ic.pdf}
    \end{subfigure}
    \hfill
    \begin{subfigure}{0.49\textwidth}
        \centering
        \includegraphics[trim=0.3cm 0.2cm 1cm 0.6cm,
    clip, width=\linewidth, height=0.35\textheight, keepaspectratio]{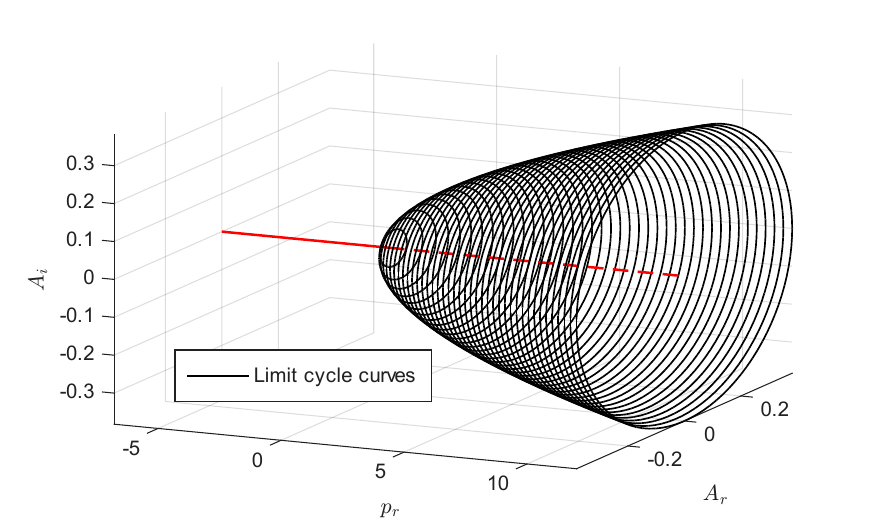}
        \caption{}
        \label{fig:hopfbifurcation15v0.5k0.65Ic-1000RtRF.pdf}
    \end{subfigure}
    
    \caption{(a) Temporal evolution of the amplitude $|A|$ with slow time $s$ for different Taylor numbers $T_a$, with fixed parameters $V = 15$, $\kappa = 1$, and $K_c = 0.51$. 
(b) Corresponding supercritical Hopf bifurcation diagrams for Taylor numbers $T_a=800$.}
    \label{fig:amp+hopf}
\end{figure*}
\begin{figure*}[t]
    \centering
    
    \begin{subfigure}{0.49\textwidth}
        \centering
        \includegraphics[trim=.1cm 0.5cm 1cm 0.6cm,
    clip, width=\linewidth, height=0.35\textheight, keepaspectratio]{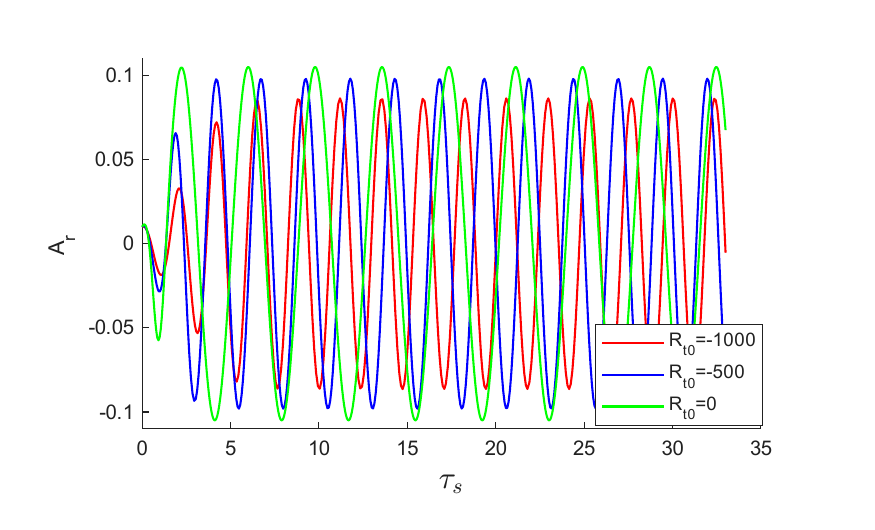}
        \caption{}
        \label{fig:Real15v0.5k0.65IcRF.pdf}
    \end{subfigure}
    \hfill
    \begin{subfigure}{0.49\textwidth}
        \centering
        \includegraphics[trim=0.1cm 0.5cm 1cm 0.6cm,
    clip, width=\linewidth, height=0.35\textheight, keepaspectratio]{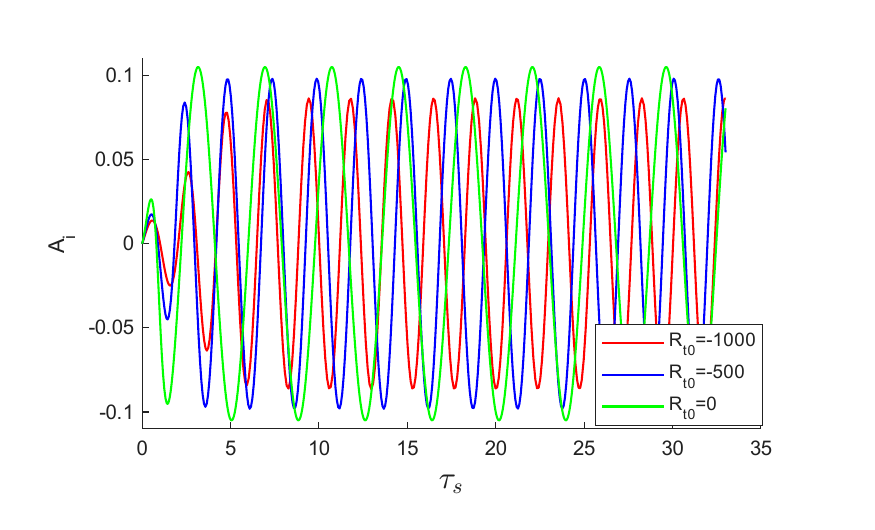}
        \caption{}
        \label{fig:Imag15v0.5k0.65IcRF.pdf}
    \end{subfigure}
    
    \caption{(a) Temporal evolution of the amplitude $|A|$ with slow time $s$ for different Taylor numbers $T_a$, with fixed parameters $V = 15$, $\kappa = 1$, and $K_c = 0.51$. 
(b) Corresponding supercritical Hopf bifurcation diagrams for Taylor numbers $T_a=800$.}
    \label{fig:Real+Imag}
\end{figure*}
\begin{figure*}[b]
	\begin{center}
			\includegraphics[trim=10cm 0.2cm 11cm 0.6cm,
    clip, width=0.9\textwidth, height=0.45\textheight]{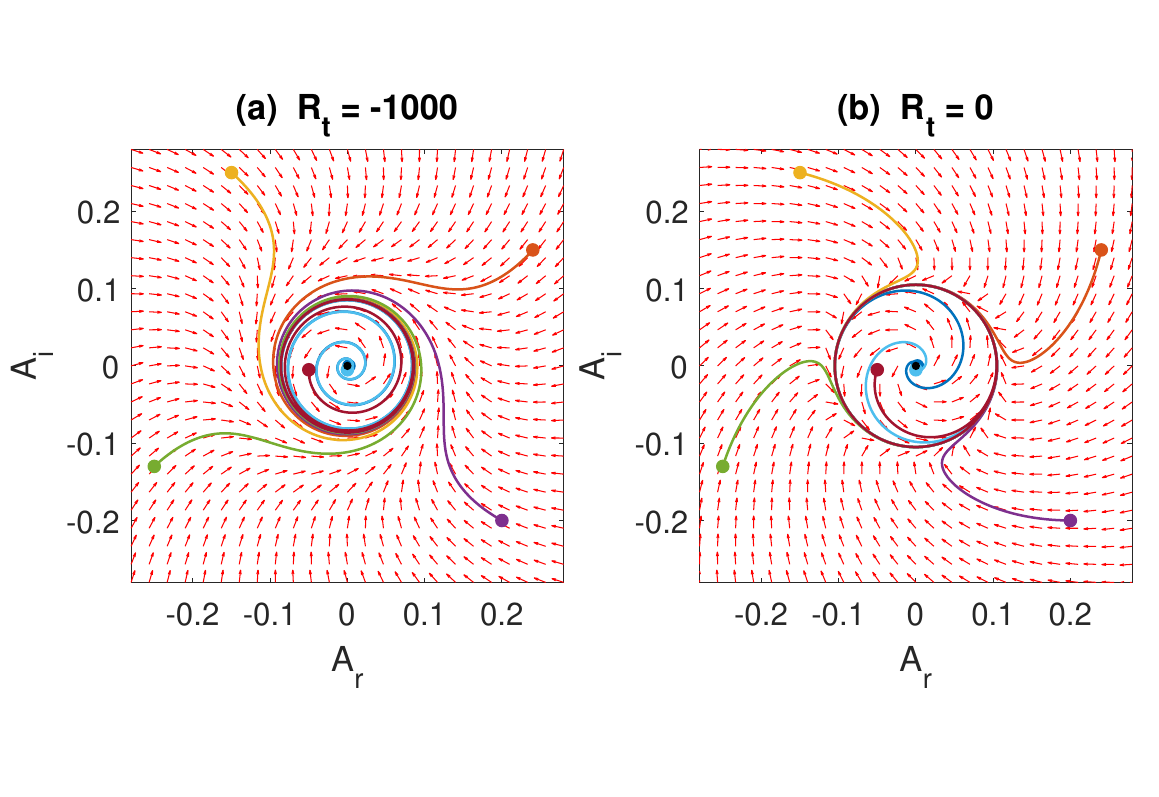}	
            \caption{(a) Steady-state distribution of cell concentration $n_0$ versus the vertical coordinate $z$. 
(b) Neutral stability curves of the Rayleigh number $R_n$ versus wavenumber $m$ for different Taylor numbers $T_a$, with $V = 15$, $K_c = 0.51$, and $\kappa = 1$. The system assumes a no-slip condition at the bottom boundary and a stress-free condition at the top. Solid and dashed curves represent stationary and overstable branches, respectively.}
   \label{fig:Hopf_PhasePortrait_Rt_-1000and0.pdf}
		\end{center}       
\end{figure*}
\begin{figure*}[b]
	\begin{center}
			\includegraphics[ width=0.9\textwidth, height=0.45\textheight,trim=1cm 6cm 1cm 6cm,
    clip]{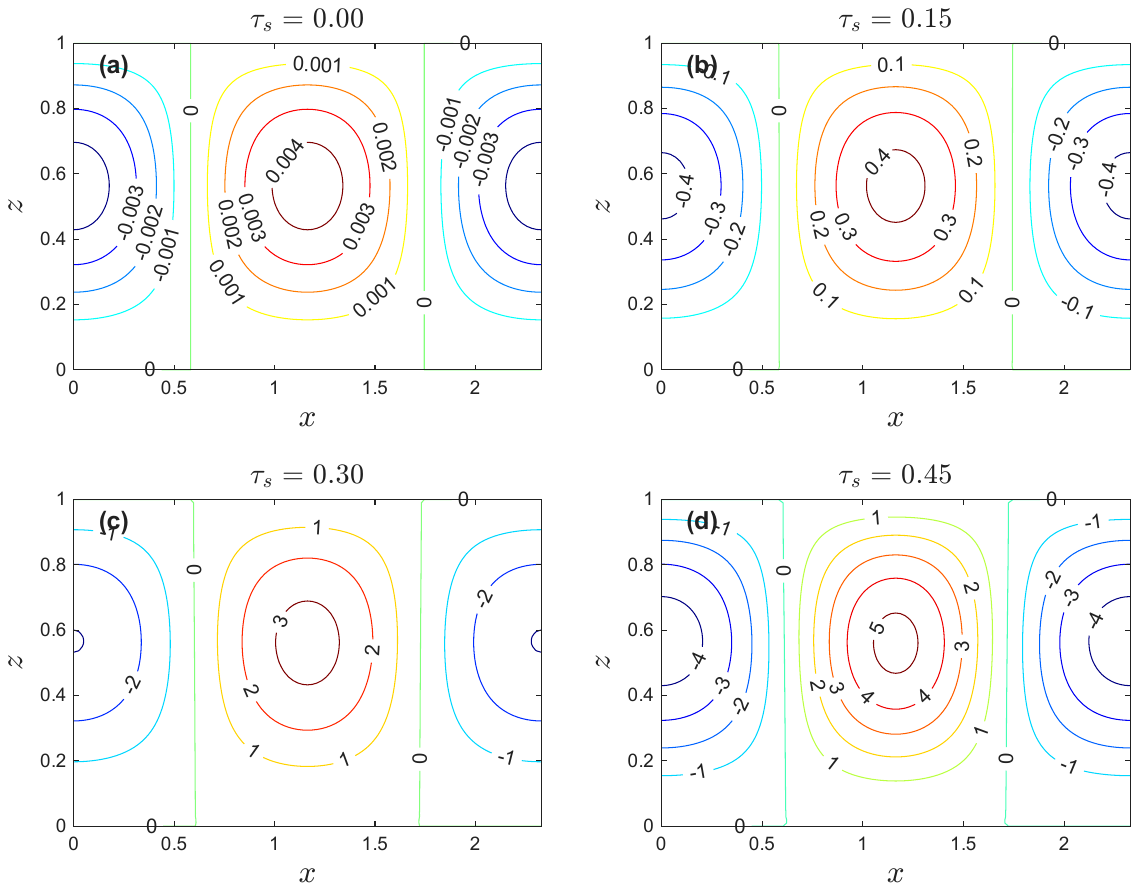}	
            \caption{Velocity field contours for $K_c = 0.5$, $V = 10$, and $\kappa = 1$ at slow times (a) $s = 0$, (b) $s = 0.04$, (c) $s = 0.08$, and (d) $s = 0.15$.}
   \label{fig:contour15v0.5k0.65Ic1100.5RtVelocity.pdf}
		\end{center}       
\end{figure*}
\begin{figure*}[b]
	\begin{center}
			\includegraphics[ width=0.9\textwidth, height=0.45\textheight,trim=1cm 6cm 1cm 6cm,
    clip]{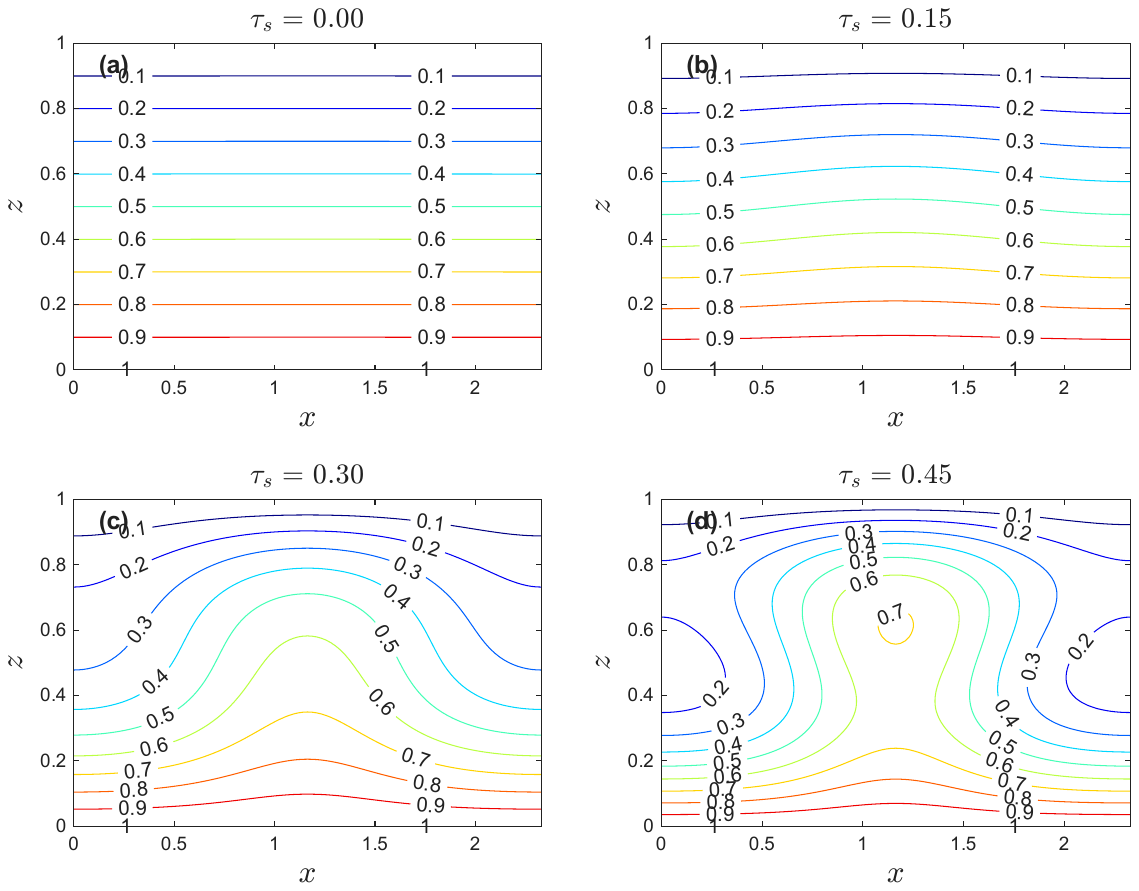}	
            \caption{Vorticity field contours for $K_c = 0.5$, $V = 10$, and $\kappa = 1$ at slow times (a) $s = 0$, (b) $s = 0.04$, (c) $s = 0.08$, and (d) $s = 0.15$.}
   \label{fig:contour15v0.5k0.65Ic1100.5RtTemp.pdf}
		\end{center}       
\end{figure*}
\begin{figure*}[b]
	\begin{center}
			\includegraphics[ width=0.9\textwidth, height=0.45\textheight,trim=1cm 6cm 1cm 6cm,
    clip]{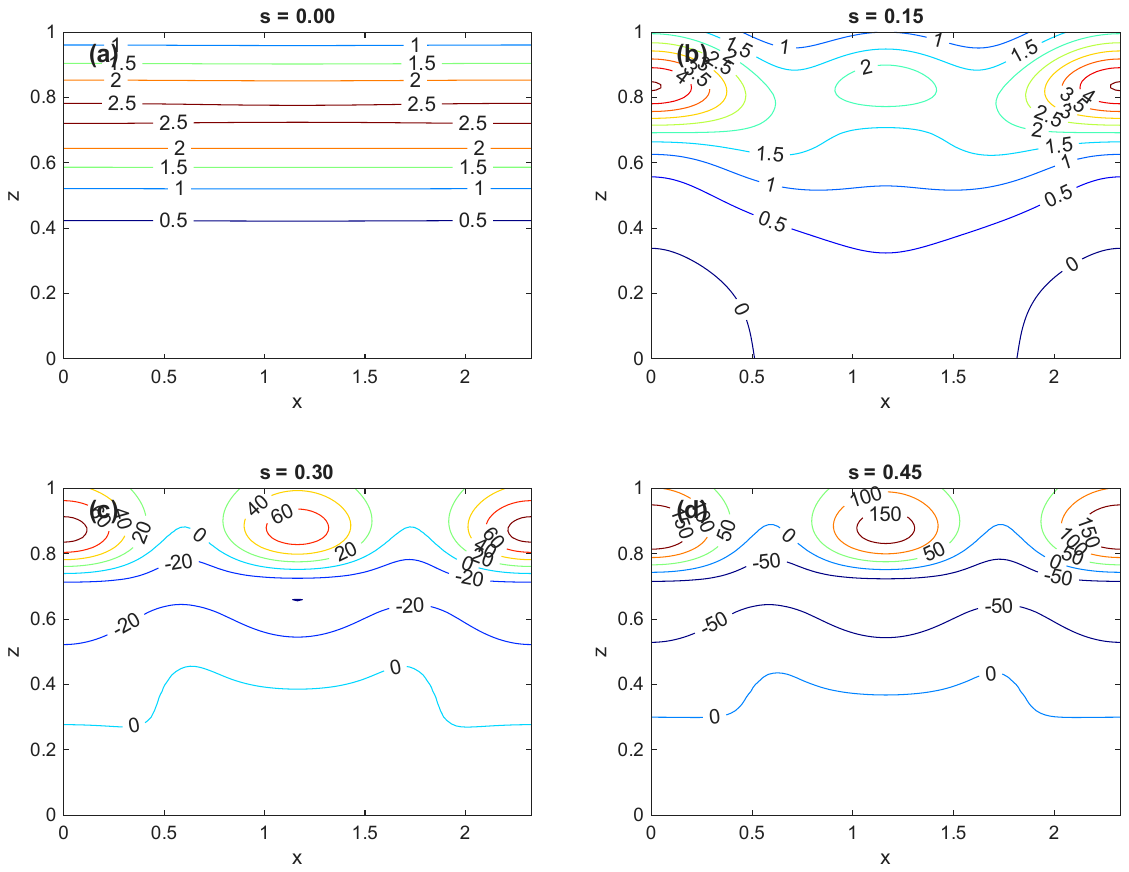}	
            \caption{cell concentration field for $K_c = 0.5$, $V = 10$, and $\kappa = 1$ at slow times (a) $s = 0$, (b) $s = 0.04$, (c) $s = 0.08$, and (d) $s = 0.15$.}
   \label{fig:contour15v0.5k0.65Ic1100.5RtConc.pdf}
		\end{center}       
\end{figure*}

\begin{figure*}[b]
	\begin{center}
			\includegraphics[ width=0.9\textwidth, height=0.35\textheight]{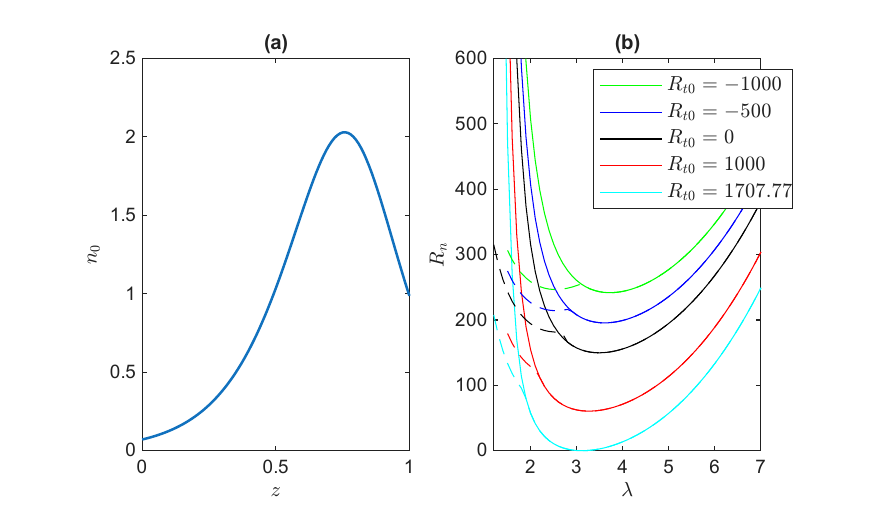}	
            \caption{(a) Basic-state microorganism concentration profile for $V=10$, $K_c=0.66$, and $\kappa=0.5$. (b) Neutral stability curves for various values of the thermal Rayleigh number $R_{t0}$ under rigid-rigid boundary conditions.}
   \label{fig:10v0.5k0.66IcRR.pdf}
		\end{center}       
\end{figure*}

\begin{figure*}[t]
    \centering
    
    \begin{subfigure}{0.49\textwidth}
        \centering
        \includegraphics[trim=1cm 0.5cm 1cm 0.6cm,
    clip, width=\linewidth, height=0.35\textheight, keepaspectratio]{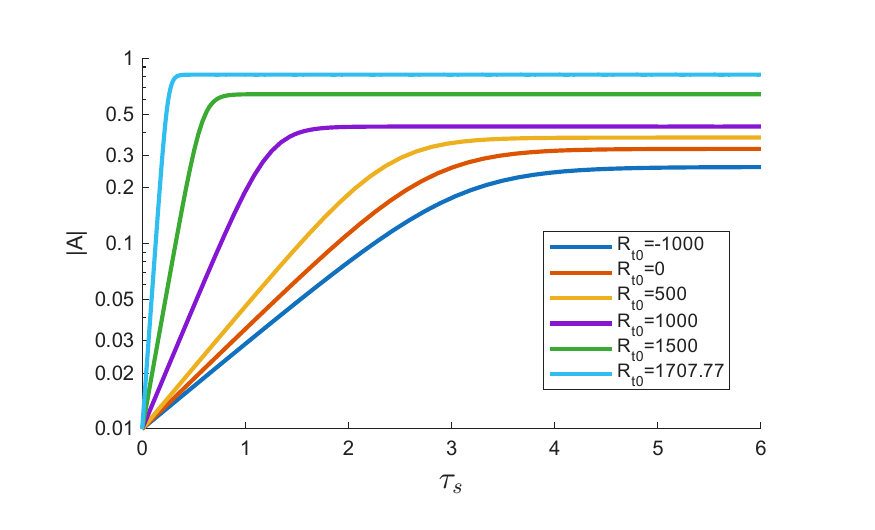}
        \caption{}
        \label{fig:Amplitude10v0.5k0.66IcRR.pdf}
    \end{subfigure}
    \hfill
    \begin{subfigure}{0.49\textwidth}
        \centering
        \includegraphics[trim=0.5cm 0.2cm 1cm 0.3cm,
    clip, width=\linewidth, height=0.35\textheight, keepaspectratio]{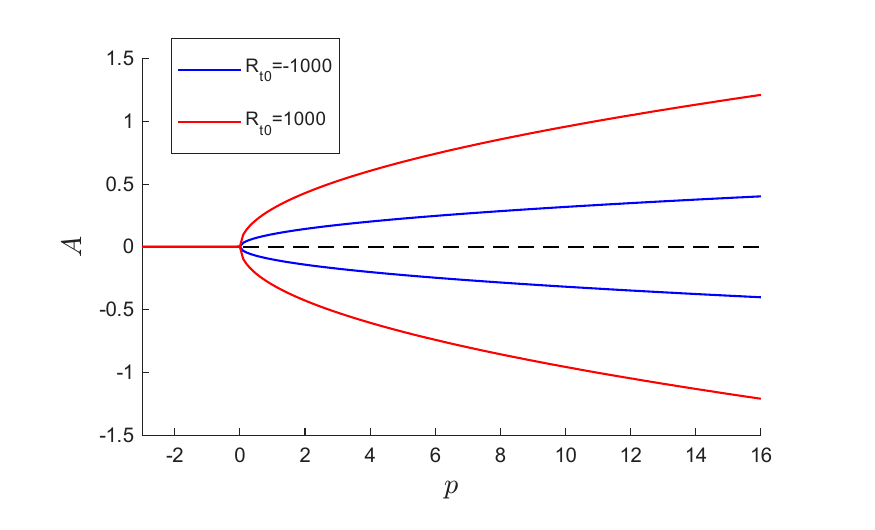}
        \caption{}
        \label{fig:bifurcation10v0.5k0.66IcRR.pdf}
    \end{subfigure}
    
    \caption{(a) Temporal evolution of the amplitude $|A|$ with slow time $s$ for different Taylor numbers $T_a$, with fixed parameters $V = 15$, $\kappa = 1$, and $K_c = 0.51$. 
(b) Corresponding supercritical Hopf bifurcation diagrams for Taylor numbers $T_a=800$.}
    \label{fig:amp+pitchfork2}
\end{figure*}

\section{Second order problem}
\label{sec5}

At \(O(\chi^2)\), nonlinear interactions between the first-order perturbations contribute to the governing equations. The second-order problem is therefore governed by
\begin{align}
\label{eqn:equation29}
&\nabla^4 w_2
-\frac{1}{P_r}\frac{\partial}{\partial \tau_f}
\left(\nabla^2 w_2\right)
-R_0\left(
\partial_{xx}
+\partial_{yy}
\right)n_2
+R_{t0}\left(
\partial_{xx}
+\partial_{yy}
\right)\theta_2
\nonumber\\
&=
-\frac{1}{P_r}
\left[
\nabla\bigl(\nabla\cdot(\boldsymbol{u}_1\cdot\nabla\boldsymbol{u}_1)\bigr)
-\nabla^2(\boldsymbol{u}_1\cdot\nabla\boldsymbol{u}_1)
\right]\cdot\hat{k},
\\
\label{eqn:equation30}
&
L_n \frac{\partial n_2}{\partial \tau_f}
+L_n \frac{dn_0}{dz}w_2
+V\nabla\cdot
\left(
n_2\hat{p}_0+n_0\hat{p}_2
\right)
-\nabla^2 n_2=-\left[L_n
\nabla\cdot(n_1\boldsymbol{u}_1)
+
V\nabla\cdot(n_1\hat{p}_1)
\right],
\\
&
\nabla^2\theta_2
-\frac{\partial\theta_2}{\partial\tau_f}
+w_2
=
\boldsymbol{u}_1\cdot\nabla\theta_1.
\end{align}

The corresponding boundary conditions at second order are
\[
w_2=\theta_2=0,
\qquad z=0,1,
\]
together with
\[
\left[
V\left(
n_2\hat{p}_0
+n_1\hat{p}_1
+n_0\hat{p}_2
\right)
-\nabla n_2
\right]\cdot\hat{k}
=0,
\qquad z=0,1.
\]

For rigid boundaries, 
\[
\frac{\partial w_2}{\partial z}=0,
\qquad z=0,1,
\]
whereas for a stress-free upper boundary,
\[
\frac{\partial^2 w_2}{\partial z^2}=0,
\qquad z=1.
\]

To simplify the subsequent analysis, the transformed variable $n_2'(x,y,z,t)=\int_z^1 n_2(x,y,z',t)\,dz'$ is introduced. In terms of this variable, the second-order system can be written as
\[
\mathcal{L}\boldsymbol{X}_2
=
\boldsymbol{f}_2.
\]

Equivalently,
\begin{equation}
\label{eqn:equation31}
\begin{aligned}
&
\begin{bmatrix}
\mathcal{L}_{11}
&
\mathcal{L}_{12}
&
\mathcal{L}_{13}
\\
-L_n\dfrac{\partial n_0}{\partial z}
&
L_n\dfrac{\partial^2}{\partial\tau_f\partial z}
-\digamma
&
0
\\
1
&
0
&
\nabla^2
-\dfrac{\partial}{\partial\tau_f}
\end{bmatrix}
\begin{bmatrix}
w_2\\
n_2'\\
\theta_2
\end{bmatrix}
=
\begin{bmatrix}
f_{21}\\
f_{22}\\
f_{23}
\end{bmatrix}.
\end{aligned}
\end{equation}

Here, $\boldsymbol{X}_2
=
\begin{bmatrix}
w_2 &
n_2' &
\theta_2
\end{bmatrix}^{T}$, and $\boldsymbol{f}_2
=
\begin{bmatrix}
f_{21} &
f_{22} &
f_{23}
\end{bmatrix}^{T}$.

The forcing vector \(\boldsymbol{f}_2\) comprises the quadratic interaction terms generated by the first-order solutions.


\section{Third order problem}
\label{sec6}

At the third order, \(O(\chi^3)\), the governing equations contain contributions associated with both the fast and slow time scales, together with the nonlinear interactions involving the first and second order perturbation fields. The resulting system is
\begin{align}
\label{eqn:equation32}
&\nabla^4 w_3
-\frac{1}{P_r}\frac{\partial}{\partial \tau_f}
\left(\nabla^2 w_3\right)
-R_0\left(
\partial_{xx}
+\partial_{yy}
\right)n_3
+R_{t0}
\left(
\partial_{xx}
+\partial_{yy}
\right)\theta_3
\nonumber\\
&=
\frac{1}{P_r}
\frac{\partial}{\partial\tau_s}
\left(\nabla^2w_1\right)
+
R_2
\left(
\partial_{xx}
+\partial_{yy}
\right)n_1
-
R_{t2}
\left(
\partial_{xx}
+\partial_{yy}
\right)\theta_1
\nonumber\\
&\quad
-\frac{1}{P_r}
\Big[
\nabla\bigl(\nabla\cdot(\boldsymbol{u}_1\cdot\nabla\boldsymbol{u}_2)\bigr)
-\nabla^2(\boldsymbol{u}_1\cdot\nabla\boldsymbol{u}_2)
+\nabla\bigl(\nabla\cdot(\boldsymbol{u}_2\cdot\nabla\boldsymbol{u}_1)\bigr)
-\nabla^2(\boldsymbol{u}_2\cdot\nabla\boldsymbol{u}_1)
\Big]\cdot\hat{k},
\\
\label{eqn:equation33}
&
L_n\frac{\partial n_3}{\partial\tau_f}
+
L_n\frac{\partial n_1}{\partial\tau_s}
+L_n\nabla\cdot
\left(
n_0\boldsymbol{u}_3
+n_1\boldsymbol{u}_2
+n_2\boldsymbol{u}_1
\right)
+
V\nabla\cdot
\left(
n_3\hat{p}_0
+n_2\hat{p}_1
+n_1\hat{p}_2
+n_0\hat{p}_3
\right)
-\nabla^2 n_3
=0,
\\
&
\nabla^2\theta_3
-\frac{\partial\theta_3}{\partial\tau_f}
+w_3
=
\frac{\partial\theta_1}{\partial\tau_s}
+\boldsymbol{u}_1\cdot\nabla\theta_2
+\boldsymbol{u}_2\cdot\nabla\theta_1.
\end{align}

The corresponding third-order boundary conditions are
\[
w_3=\theta_3=0,
\qquad z=0,1,
\]
\[
\left[
V\left(
n_3\hat{p}_0
+n_2\hat{p}_1
+n_1\hat{p}_2
+n_0\hat{p}_3
\right)
-\nabla n_3
\right]\cdot\hat{k}
=0,
\qquad z=0,1.
\]

For rigid boundaries, 
\[
\frac{\partial w_3}{\partial z}=0,
\qquad z=0,1,
\]
whereas for a stress-free upper boundary,
\[
\frac{\partial^2w_3}{\partial z^2}=0,
\qquad z=1.
\]

To evaluate the higher-order contributions associated with phototactic swimming, the swimming-direction vector is expanded as
\begin{align*}
\hat{p}
&=
\hat{p}_0
+\chi\hat{p}_1
+\chi^2\hat{p}_2
+\chi^3\hat{p}_3
+\cdots
=
T\!\left(
K_0
+\chi K_1
+\chi^2K_2
+\chi^3K_3
+\cdots
\right)\hat{k}.
\end{align*}

The individual terms in the expansion are
\begin{align*}
\hat{p}_0
&=
T(K_0)\hat{k}, \qquad \hat{p}_1=K_1\frac{dT}{dK_0}\hat{k},
\quad
\hat{p}_2
=
\left(
K_2\frac{dT}{dK_0}
+\frac{1}{2}
\frac{d^2T}{dK_0^2}
K_1^2
\right)\hat{k},
\\
\hat{p}_3
&=
\left(
K_3\frac{dT}{dK_0}
+\frac{d^2T}{dK_0^2}K_1K_2
+\frac{1}{6}
\frac{d^3T}{dK_0^3}
K_1^3
\right)\hat{k}.
\end{align*}

The corresponding light intensity coefficients are given by
\begin{align*}
K_0
&=
J_s
e^{\kappa\int_1^z n_0(z')\,dz'},
\quad
K_1
=
K_0\kappa
\int_1^z
n_1(x,y,z',t)\,dz',
\\
K_2
&=
K_0
\Bigg[
\kappa
\int_1^z
n_2(x,y,z',t)\,dz'
+\frac{1}{2}\kappa^2
\left(
\int_1^z
n_1(x,y,z',t)\,dz'
\right)^2
\Bigg],
\\
K_3
&=
K_0
\Bigg[
\kappa
\int_1^z
n_3(x,y,z',t)\,dz'
+\kappa^2
\int_1^z
n_1(x,y,z',t)\,dz'
\int_1^z
n_2(x,y,z',t)\,dz'
\\
&\qquad
+\frac{1}{6}\kappa^3
\left(
\int_1^z
n_1(x,y,z',t)\,dz'
\right)^3
\Bigg].
\end{align*}

Introducing the transformed variable $n_3'(x,y,z,t)=\int_z^1 n_3(x,y,z',t)\,dz'$, the third-order problem may be written in the operator form as
\[
\mathcal{L}\boldsymbol{X}_3
=
\boldsymbol{f}_3.
\]

Equivalently,
\begin{equation}
\label{eqn:equation34}
\begin{aligned}
&
\begin{bmatrix}
\mathcal{L}_{11}
&
\mathcal{L}_{12}
&
\mathcal{L}_{13}
\\
-L_n\dfrac{\partial n_0}{\partial z}
&
L_n\dfrac{\partial^2}{\partial\tau_f\partial z}
-\digamma
&
0
\\
1
&
0
&
\nabla^2
-\dfrac{\partial}{\partial\tau_f}
\end{bmatrix}
\begin{bmatrix}
w_3\\
n_3'\\
\theta_3
\end{bmatrix}
=
\begin{bmatrix}
f_{31}\\
f_{32}\\
f_{33}
\end{bmatrix}.
\end{aligned}
\end{equation}

Here, $\boldsymbol{X}_3
=
\begin{bmatrix}
w_3 &
n_3' &
\theta_3
\end{bmatrix}^{T}$, $\boldsymbol{f}_3
=
\begin{bmatrix}
f_{31} &
f_{32} &
f_{33}
\end{bmatrix}^{T}$. The vector \(\boldsymbol{f}_3\) collects all forcing terms generated by the lower-order solutions, including slow-time contributions and quadratic interactions between first-  and second-order fields. These terms provide the solvability condition from which the amplitude equation is subsequently derived. At the critical state, the linear operator \(\mathcal{L}\) possesses a non-trivial null space and therefore remains singular. Consequently, the higher-order inhomogeneous problem admits a solution only when the forcing terms satisfy an appropriate solvability condition. This condition is obtained by introducing the adjoint problem. Let \(\langle \cdot,\cdot\rangle\) the inner product and \(\mathcal{L}^{*}\) be the adjoint operator be defined through $\langle \mathcal{L}^{*}\boldsymbol{X}^{*},\boldsymbol{X}_{1}\rangle
=\langle \boldsymbol{X}^{*},\mathcal{L}\boldsymbol{X}_{1}\rangle$, for arbitrary vectors \(\boldsymbol{X}_{1}\) and \(\boldsymbol{X}^{*}\). According to the Fredholm alternative theorem \cite{skene2024weakly}, the higher-order system (\ref{eqn:equation34}) is solvable only if its forcing vector $\boldsymbol{f}_3$ is orthogonal to the null space of the adjoint operator. Thus, the solvability condition takes the form
\begin{equation}
\label{eqn:equation40}
\langle \boldsymbol{X}^{*},\boldsymbol{f}_{3}\rangle =0.
\end{equation}

Evaluation of Equation ~(\ref{eqn:equation40}) and subsequent simplification yield the Stuart-Landau equation governing the slow time evolution of the disturbance amplitude 
\begin{equation}
\label{eqn:equation41}
\frac{dA}{d\tau_s}
=
pA-l|A|^{2}A.
\end{equation}

The complex coefficient \(p=p_r+ip_i\) characterizes the linear growth, while \(l=l_r+il_i\) denotes the Landau coefficient. Expressing the amplitude in polar form as $A(\tau_s)=|A(\tau_s)|e^{i\beta}$, the Stuart-Landau equation separates into the amplitude and phase equations
\begin{align}
\label{eqn:equation42}
\frac{d|A|^2}{d\tau_s}
&=
2p_r|A|^2
-
2l_r|A|^4,
\\
\label{eqn:equation43}
\frac{d\beta}{d\tau_s}
&=
p_i
-
l_i|A|^2.
\end{align}

In the case \(p_i=l_i=0\), the phase remains constant, and the dynamics are entirely determined by the amplitude equation. The resulting instability is stationary and takes the form of a pitchfork bifurcation. The nature of this bifurcation depends on the sign of \(l_r\): \(l_r>0\) corresponds to a supercritical bifurcation, whereas \(l_r<0\) corresponds to a subcritical bifurcation. When \(p_i\) and/or \(l_i\) are nonzero, the phase exhibits temporal variation, giving rise to oscillatory solutions characteristic of a Hopf bifurcation.

\section{Numerical results}
\label{sec7}

The equilibrium solution described by Eqs.~(\ref{basicstate})-(\ref{basicstatebc}) and the generalized eigenvalue problem given by Eq.~(\ref{eqn:equation24}), together with the boundary conditions (\ref{eqn:equation25})-(\ref{eqn:equation28}), are computed numerically using MATLAB's \texttt{bvp4c} boundary-value solver. The same numerical procedure is employed to obtain the solutions of the second-order and adjoint systems. Subsequently, the Stuart-Landau equation (\ref{eqn:equation41}) is integrated with MATLAB's \texttt{ode45} routine, subject to the initial condition \(A(0)=A_0\), where \(A_0\) denotes the initial disturbance amplitude. The parameter set employed in the present computations corresponds to a phototactic suspension of \textit{Chlamydomonas} and follows the ranges reported in earlier studies \citep{kumar2023effect,ref9}. The mean cell concentration is fixed at \(\bar n=10^{6}\,\mathrm{cm^{-3}}\), while the relative density difference is taken as \(\Delta\rho/\rho=5\times10^{-2}\). The cell volume and radius are specified as \(5\times10^{-10}\,\mathrm{cm^3}\) and \(10^{-3}\,\mathrm{cm}\), respectively. Furthermore, the microorganism diffusivity is chosen as \(D=5\times10^{-4}\,\mathrm{cm^2/s}\), thermal diffusivity is $\beta_f=2\times10^{-3}\,\mathrm{cm^2/s}$, whereas the average swimming speed is \(U=10^{-2}\,\mathrm{cm/s}\). The scaled swimming velocity parameter is prescribed by \(V=20L\). The incident light intensity, Lewis number, and Prandtl number are maintained at \(J_s=0.8\), $L_n=4$, and \(P_r=5\), respectively. The phototactic parameter \(\Upsilon\) is related to the critical light intensity \(K_c\), and the range \(-1.1\leq \Upsilon \leq 1.1\) corresponds to \(0.3\leq K_c\leq0.8\) \citep{ref17}.

Figure~\ref{10v1.00.5Iclinear+basic.pdf}(a) illustrates the steady-state distribution of microorganisms for $V=10$, $K_c=0.5$, and $\kappa=1$. The concentration profile exhibits a pronounced maximum around $z\approx0.6$, indicating the formation of a dense sublayer of algae cells within the suspension. The corresponding neutral stability curves for several values of the thermal Rayleigh number $R_{t0}$ ranging from $-1000$ to $1100.5$ are presented in Figure~\ref{10v1.00.5Iclinear+basic.pdf}(b) for a rigid lower boundary and a stress-free upper surface. It is evident that negative values of $R_{t0}$ (cooling from below) shift the neutral curves upward, thereby increasing the critical bioconvection Rayleigh number and delaying the onset of instability. This stabilizing effect arises because the imposed thermal stratification suppresses buoyancy-driven motion and weakens convective transport within the suspension. In contrast, positive values of $R_{t0}$ (heating from below) promote thermal convection, leading to a reduction in the critical bioconvection Rayleigh number. As $R_{t0}$ increases and approaches $1100.5$, the critical bioconvection Rayleigh number $R_0$ decreases significantly and eventually tends toward zero. The thermal buoyancy generated by bottom heating reinforces the phototaxis-induced cell accumulation, thereby enhancing convective motion and accelerating the onset of bioconvection. Therefore, increasing the thermal Rayleigh number destabilizes the suspension.

Figure~\ref{fig:linearsol.pdf} depicts the eigenfunctions corresponding to the vertical velocity $W_1(z)$, temperature perturbation $\Theta_1(z)$, and microorganism concentration perturbation $N_1(z)$ for $V=10$, $K_c=0.5$, and $\kappa=1$. The vertical velocity eigenfunction shown in Figure~\ref{fig:linearsol.pdf}(a) remains negative throughout the layer, indicating a downward fluid motion associated with the convective cell. As the thermal Rayleigh number $R_{t0}$ increases, the magnitude of $W_1$ becomes larger, reflecting the strengthening influence of thermal buoyancy on the flow. The corresponding temperature perturbation eigenfunction $\Theta_1$ is presented in Figure~\ref{fig:linearsol.pdf}(b). Although $\Theta_1$ satisfies homogeneous boundary conditions at both surfaces, its non-zero distribution arises from its coupling with the velocity field through the energy equation. Figure~\ref{fig:linearsol.pdf}(c) shows the microorganism concentration perturbation eigenfunction $N_1$. An increase in $R_{t0}$ enhances the magnitude of the temperature perturbation, indicating a stronger thermal response within the suspension. Positive $N_1$ values in the lower section indicate an accumulation of microorganisms, creating a zone of negative buoyancy that drives downward fluid flow. Conversely, negative $N_1$ values near the upper surface reflect a drop in cell concentration, yielding a positively buoyant region that resists this descending convective motion.

\begin{table*}[h!]
\centering
\begin{tabular}{|c|c|c|c|c|c|c|}
\hline\hline
$R_t$ & $-1000$ & $-500$ & $0$ & $500$ & $1000$ & $1100.5$ \\
\hline
$m_c$ & 3.5 & 3.1 & 2.9 & 2.8 & 2.6 & 2.7 \\
\hline
$R_0$ & 237.41 & 187.2 & 133.98 & 77.67 & 15.75 & 0 \\
\hline
$p$ &
$2.3288$ &
$2.6406$ &
$3.8118$ &
$7.6277$ &
$19.4549$ &
$15.2251$ \\
\hline
$l$ &
$765.0193$ &
$618.5176$ &
$626.7375$ &
$749.9287$ &
$366.7513$ &
$166.8236$ \\
\hline
$\sqrt{p_r/l_r}$ &
$0.0552$ &
$0.0653$ &
$0.0780$ &
$0.1009$ &
$0.2303$ &
$0.3021$ \\
\hline
$l_{A\bar{A}}$ &
$773.5782$ &
$633.4999$ &
$651.5716$ &
$791.8667$ &
$379.9602$ &
$-5.6957$ \\
\hline
$l_{AA}$ &
$-8.5589$ &
$-14.9823$ &
$-24.8341$ &
$-41.9379$ &
$-13.2089$ &
$172.5193$ \\
\hline
$l_u$ &
$0.037995$ &
$0.22324$ &
$0.3913$ &
$0.96414$ &
$2.3717$ &
$0.93153$ \\
\hline
$l_n$ &
$-7.5492$ &
$-14.5214$ &
$-25.2254$ &
$-45.4877$ &
$-34.2129$ &
$-2.8464$ \\
\hline
$l_t$ &
$-1.0477$ &
$-0.68418$ &
$0$ &
$2.5856$ &
$18.6323$ &
$174.4341$ \\
\hline\hline
\end{tabular}
\caption{Weakly nonlinear coefficients at the critical wavenumber and critical bioconvection Rayleigh number for various thermal Rayleigh numbers $R_t$, with $V = 10$, $K_c = 0.5$, and both boundaries rigid.}
\label{tab:1}
\end{table*}

\begin{table*}[h!]
\centering
\begin{tabular}{|c|c|c|c|c|c|c|}
\hline\hline
$R_t$ & $-1000$ & $-500$ & $0$ & $500$ & $1000$ & $1100.5$ \\
\hline
$m_c$ & 2.2 & 2.2 & 2.1 & 3.1 & 2.8 & 2.7 \\
\hline
$R_0$ & 138.12 & 114.54 & 90.55 & 57.76 & 9.93 & 0 \\
\hline
$p$ &
$0.612+3.1090i$ &
$1.1668+3.5728i$ &
$2.2327+4.1303i$ &
$2.083$ &
$5.8541$ &
$15.0805$ \\
\hline
$l$ &
$82.2821+60.0753i$ &
$121.8360+112.9758i$ &
$202.6080+224.1402i$ &
$60.2376$ &
$123.4810$ &
$12.5296$ \\
\hline
$\sqrt{p_r/l_r}$ &
$0.0862$ &
$0.0979$ &
$0.1050$ &
$0.1860$ &
$0.2177$ &
$1.0971$ \\
\hline
$l_{A\bar{A}}$ &
$45.4144+39.917i$ &
$68.5412+66.4711i$ &
$101.5827+116.2744i$ &
$27.0468$ &
$84.1008$ &
$0.77965$ \\
\hline
$l_{AA}$ &
$36.8677+20.1584i$ &
$53.2947+46.5047i$ &
$101.0253+107.8658i$ &
$33.1908$ &
$39.3802$ &
$11.7499$ \\
\hline
$l_u$ &
$-0.4217+0.2501i$ &
$-2.0567+1.7128i$ &
$-3.658+3.5824i$ &
$-0.1315$ &
$-0.0524$ &
$0.0315$ \\
\hline
$l_n$ &
$38.086+18.89i$ &
$55.5502+43.9460i$ &
$104.6833+104.2834i$ &
$33.2492$ &
$39.1419$ &
$2.1588$ \\
\hline
$l_t$ &
$-0.7966+1.0182i$ &
$-0.1987+0.8458i$ &
$0$ &
$0.073$ &
$0.2907$ &
$9.5596$ \\
\hline\hline
\end{tabular}
\caption{Weakly nonlinear coefficients at the critical wavenumber and critical bioconvection Rayleigh number for various thermal Rayleigh numbers $R_t$, with $V = 15$, $K_c = 0.65$, and both boundaries rigid.}
\label{tab:2}
\end{table*}

\begin{table*}[h!]
\centering
\begin{tabular}{|c|c|c|c|c|c|c|c|}
\hline\hline
$R_t$ & $-1000$ & $-500$ & $0$ & $500$ & $1000$ & $1500$ & $1707.77$ \\
\hline
$m_c$ & 3.7 & 3.6 & 3.5 & 3.4 & 3.3 & 3.2 & 3.1 \\
\hline
$R_0$ & 241.77 & 195.53 & 149.8 & 104.74 & 60.59 & 17.54 & 0 \\
\hline
$p$ &
$7.8747$ &
$8.3749$ &
$1.5287$ &
$2.0150$ &
$3.0633$ &
$7.1516$ &
$17.7205$ \\
\hline
$l$ &
$98.9736$ &
$104.6115$ &
$10.9171$ &
$12.8345$ &
$16.6294$ &
$17.3847$ &
$26.4624$ \\
\hline
$\sqrt{p_r/l_r}$ &
$0.2821$ &
$0.2829$ &
$0.3742$ &
$0.3962$ &
$0.4292$ &
$0.6414$ &
$0.8183$ \\
\hline
$l_{A\bar{A}}$ &
$94.2594$ &
$98.5338$ &
$1.2044$ &
$2.6518$ &
$6.3474$ &
$9.0882$ &
$0.016847$ \\
\hline
$l_{AA}$ &
$4.7142$ &
$6.0776$ &
$9.7127$ &
$10.1827$ &
$10.2820$ &
$8.2964$ &
$26.4456$ \\
\hline
$l_u$ &
$-0.0557$ &
$-0.0495$ &
$-0.0622$ &
$-0.0594$ &
$-0.0534$ &
$-0.0365$ &
$-0.0449$ \\
\hline
$l_n$ &
$5.3299$ &
$6.4079$ &
$9.7750$ &
$10.1867$ &
$10.1676$ &
$7.7824$ &
$0.1647$ \\
\hline
$l_t$ &
$-0.5599$ &
$-0.2808$ &
$0$ &
$0.05539$ &
$0.16785$ &
$0.55055$ &
$26.3258$ \\
\hline\hline
\end{tabular}
\caption{Weakly nonlinear coefficients at the critical wavenumber and critical bioconvection Rayleigh number for various thermal Rayleigh numbers $R_t$, with $V = 10$, $K_c = 0.66$, and both boundaries rigid.}
\label{tab:3}
\end{table*}

Figure~\ref{fig:adjsol.pdf} depicts the adjoint eigenfunctions $W^{*}$, $\Theta^{*}$, and $N^{*}$ corresponding to the linear stability problem. The magnitude of $W^{*}$ increases with increasing thermal Rayleigh number $R_{t0}$, indicating a stronger response of the adjoint velocity field to thermal forcing. The adjoint temperature eigenfunction $\Theta^{*}$ undergoes a pronounced change in amplitude and sign, while the adjoint concentration eigenfunction $N^{*}$ preserves its overall shape but exhibits a gradual reduction in amplitude as $R_{t0}$ increases. After determining the critical Rayleigh number and the corresponding critical wavenumber from the linear stability analysis, the weakly nonlinear procedure is continued by solving the second-order resolvent system (\ref{eqn:equation31}). The resulting second-order solutions are subsequently employed in the evaluation of the Landau saturation coefficient $l$. The quadratic self-interaction of the fundamental mode, described by Equations~(\ref{eqn:equation22})-(\ref{eqn:equation231}), generates a first-harmonic contribution of the form
\[
A^2
\begin{bmatrix}
W_{22}\\
N_{22}
\end{bmatrix}
e^{2i(\lambda_1x+\lambda_2y)+2\eta\tau_f},
\]
whose corresponding eigenfunctions are displayed in Figure~\ref{fig:secorder2sol.pdf}. In addition to the first harmonic, the nonlinear interactions produce basic-state corrections to the temperature and microorganism concentration fields, represented by $\Theta_{20}|A|^2$ and $N_{20}|A|^2$, respectively. The influence of the thermal Rayleigh number $R_{t0}$ on these second-order corrections for $R_{t0}=-500$, $0$, and $500$ is illustrated in Figure~\ref{fig:secorder20sol.pdf}.

Figure~\ref{fig:amp+pitchfork}(a) illustrates the temporal evolution of the amplitude $|A|$ with the slow time variable $\tau_s$ for different values of the thermal Rayleigh number $R_{t0}$. In all cases, the amplitude initially grows from a small disturbance and subsequently approaches a finite steady value, indicating the saturation of the convective mode by nonlinear effects. To quantify this behaviour, the weakly nonlinear coefficients appearing in the Stuart-Landau amplitude equation were evaluated at the critical Rayleigh number and critical wavenumber for various values of the thermal Rayleigh number $R_{t0}$, and the corresponding results are listed in Table~\ref{tab:1}. Here, $p$ denotes the linear growth-rate coefficient, while $l$ represents the nonlinear saturation coefficient. The quantity $\sqrt{p_r/l_r}$ gives the equilibrium amplitude of convection. Furthermore, $l_{A\bar{A}}$ and $l_{AA}$ correspond to the contributions arising from the mean-state correction and first-harmonic interaction, respectively, whereas $l_{\boldsymbol{u}}$, $l_n$, and $l_t$ quantify the effects of the velocity, microorganism concentration, and temperature fields on the nonlinear saturation mechanism. The thermal contribution coefficient $l_t$ remains negative for $R_{t0}<0$, becomes zero at $R_{t0}=0$, and increases substantially as $R_{t0}$ increases, attaining a value of $174.4341$ at $R_{t0}=1100.5$, which demonstrates that thermal buoyancy progressively reinforces the nonlinear convective motion.

Figure~\ref{fig:amp+pitchfork}(b) displays the bifurcation structure predicted by the Stuart-Landau amplitude equation. Since the Landau coefficient remains real and positive throughout the parameter range considered, the instability develops through a supercritical pitchfork bifurcation. For $p<0$, the trivial solution ($|A|=0$) is stable and corresponds to the quiescent basic state. As $p$ becomes positive, the basic state loses stability, and two symmetric finite-amplitude branches, $\pm\sqrt{p/l}$, emerge continuously from the bifurcation point. The nonlinear terms subsequently limit the growth of disturbances, leading to stable saturated convection. It is also evident that the amplitude of the bifurcated branches increases with increasing thermal Rayleigh number $R_{t0}$, indicating that heating from below strengthens the nonlinear convective state and enhances the intensity of bioconvection. Consequently, larger values of $R_{t0}$ lead to more vigorous finite-amplitude convection while preserving the supercritical nature of the bifurcation.

Figure~\ref{fig:vectorfield10v1k0.5Ic.pdf} presents the phase portraits associated with the Stuart-Landau amplitude equation, $\frac{dA}{d\tau_s}=pA-lA^3$, while Figure~\ref{fig:potential10v1k0.5Ic.pdf} shows the corresponding potential function, $V(A)=-\frac{p}{2}A^2+\frac{l}{4}A^4$, which satisfies the relation $\frac{dA}{d\tau_s}=-\frac{dV}{dA}$. The results are shown for three representative thermal Rayleigh numbers, namely $R_{t0}=-1000$, $R_{t0}=0$, and $R_{t0}=1000$, corresponding to cooling from below, the absence of thermal forcing, and heating from below, respectively. For $p<0$ Figure~\ref{fig:vectorfield10v1k0.5Ic.pdf}(a), with $p=-2.3288$, $-3.8118$, and $-19.4549$ for $R_{t0}=-1000$, $0$, and $1000$, respectively, all trajectories are directed toward the origin. The associated potential possesses a single minimum at $A=0$, indicating that the conductive state is the only stable equilibrium. At the critical point $p=0$ Figure~\ref{fig:vectorfield10v1k0.5Ic.pdf}(b), the linear growth term vanishes and the vector field becomes extremely weak in the vicinity of the origin. The potential reduces to a quartic well centered at $A=0$, and the system approaches equilibrium much more slowly than in the $p<0$ regime, exhibiting the well-known phenomenon of critical slowing down. For $p>0$ Figure~\ref{fig:vectorfield10v1k0.5Ic.pdf}(c), with $p=2.3288$, $3.8118$, and $19.4549$ for $R_{t0}=-1000$, $0$, and $1000$, respectively, the origin becomes unstable and trajectories evolve toward two symmetric stable states. Correspondingly, the potential assumes a double-well structure with a local maximum at $A=0$ and two minima located at $A=\pm\sqrt{p/l}$. These equilibrium amplitudes are $\pm0.0552$, $\pm0.0780$, and $\pm0.2303$ for $R_{t0}=-1000$, $0$, and $1000$, respectively.

Figure~\ref{fig:15v0.5k0.65IcRF.pdf}(a) presents the basic-state profile for the fixed parameter values \(V=15\), \(\kappa=0.5\), and \(K_c=0.65\). The corresponding neutral stability curves for different values of the thermal Rayleigh number \(R_{t0}\) are shown in Figure~\ref{fig:15v0.5k0.65IcRF.pdf}(b). In the absence of thermal heating and under cooling from below (\(R_{t0}\leq0\)), the overstable branch remains distinct from the stationary branch and provides the minimum critical Rayleigh number. For \(R_{t0}=0\), the critical values are \(R_0=90.55\) and \(a_0=2.1\), while the instability sets in through oscillatory convection with frequency \(\mathrm{Im}(\eta)=6.82\). The associated eigenvalues occur as complex-conjugate pairs, confirming the oscillatory nature of the instability. This behavior persists throughout the cooling regime (\(R_{t0}\geq-1000\)), where convection develops beneath the gravitationally unstable microorganism sublayer and is suppressed above it. Consequently, the competition between the destabilizing phototactic accumulation of microorganisms and the stabilizing density stratification gives rise to overstability \cite{ref14}. As the thermal Rayleigh number increases (\(R_{t0}>0\)), the stationary and overstable branches coexist; however, the minimum critical Rayleigh number gradually shifts from the oscillatory branch to the stationary branch, indicating a transition from oscillatory to stationary convection. With a further increase in \(R_{t0}\), the critical Rayleigh number approaches the classical Chandrasekhar limit (\(R_t\approx1100.5\)) for the rigid-free boundary configuration. This transition reflects a change in the dominant instability mechanism. At low thermal Rayleigh numbers, convection is primarily driven by the localized accumulation of phototactic microorganisms, whereas increasing thermal buoyancy destabilizes the entire fluid layer from below. The resulting thermal buoyancy progressively overwhelms the phototactic contribution, lowering the threshold for convection and causing the instability to become predominantly buoyancy-driven. Thus, the system exhibits a clear transition from phototaxis-dominated oscillatory instability at low \(R_{t0}\) to thermally driven stationary convection at sufficiently large positive \(R_{t0}\).

Figure~\ref{fig:amp+hopf}(a) shows the temporal evolution of the amplitude \(|A|\) with the slow time variable \(\tau_s\) for different values of the thermal Rayleigh number \(R_{t0}\), while keeping the remaining parameters fixed at \(V=15\), \(\kappa=0.5\), and \(K_c=0.65\). For \(R_{t0}\leq0\), the primary instability is oscillatory, whereas for \(R_{t0}>0\), it becomes stationary. In both regimes, the amplitude initially grows exponentially from a small perturbation and subsequently saturates at a finite value due to nonlinear effects, with the saturated amplitude increasing as the thermal Rayleigh number increases. The weakly nonlinear coefficients evaluated at the critical Rayleigh number and critical wavenumber are listed in Table~\ref{tab:2}. To illustrate the nonlinear dynamics in the oscillatory regime, Figure~\ref{fig:amp+hopf}(b) presents the phase portrait in the \((A_r,A_i)\) plane for \(R_{t0}=-1000\). The trajectories evolve away from the unstable trivial equilibrium at the origin and converge to a closed orbit, indicating the formation of a stable limit cycle. This behavior is characteristic of a supercritical Hopf bifurcation. The bifurcation occurs when the linear growth coefficient satisfies \(p_r=0\). For \(p_r<0\), the trivial equilibrium \((A_r,A_i)=(0,0)\) is stable, and all perturbations decay to the origin. As \(p_r>0\), the equilibrium loses stability, and a stable periodic orbit emerges with finite amplitude $|A|=\sqrt{\frac{p_r}{l_r}}$, representing the nonlinear oscillatory state.

Figures~\ref{fig:Real+Imag}(a) and \ref{fig:Real+Imag}(b) show the temporal evolution of \(\mathrm{Re}(A)\) and \(\mathrm{Im}(A)\), respectively, for \(R_{t0}=-1000\), \(-500\), and \(0\). In these cases, both components exhibit sustained bounded oscillations with constant amplitude and frequency, confirming the existence of a stable nonlinear periodic state throughout the oscillatory regime. As the thermal Rayleigh number increases beyond \(R_{t0}=0\), the primary instability changes from oscillatory to stationary, in agreement with the linear stability analysis. Consequently, the stable limit cycle associated with the supercritical Hopf bifurcation disappears, and the system evolves toward a time-independent finite-amplitude equilibrium. Thus, the nonlinear dynamics undergo a transition from a supercritical Hopf bifurcation (\(R_{t0}\leq0\)) to a supercritical pitchfork bifurcation (\(R_{t0}>0\)).

Figure~\ref{fig:Hopf_PhasePortrait_Rt_-1000and0.pdf} illustrates the phase portraits corresponding to the supercritical Hopf bifurcation for \(V=15\), \(\kappa=0.5\), and \(K_c=0.65\), with (a) \(R_{t0}=-1000\) and (b) \(R_{t0}=0\). As the control parameter crosses the critical value (\(a_r=0\)), the trivial equilibrium loses stability and a stable limit cycle emerges. The trajectories originating from different initial conditions spiral towards this limit cycle, confirming the stable oscillatory state predicted by the Landau amplitude equation. Moreover, the radius of the limit cycle increases as \(R_{t0}\) increases from \(-1000\) to \(0\), indicating a larger saturated oscillation amplitude.

Figures~\ref{fig:contour15v0.5k0.65Ic1100.5RtVelocity.pdf}-\ref{fig:contour15v0.5k0.65Ic1100.5RtConc.pdf} illustrate the evolution of the velocity, temperature, and cell concentration fields at different slow times \(\tau_s\) for \(R_{t0}=1100.5\), \(V=15\), \(K_c=0.65\), and \(\kappa=0.5\). At the initial stage (\(\tau_s=0\)), the perturbation velocity is very weak, while the isotherms remain close to the basic-state temperature profile and the cell concentration is nearly identical to the basic-state distribution, with maximum accumulation around \(z\approx0.75\). As \(\tau_s\) increases, the velocity field strengthens, the isotherms become progressively distorted, and the concentration contours deform, indicating the development of convective motion. At \(\tau_s=0.45\), the convective circulation becomes well established, accompanied by enhanced heat transport and redistribution of microorganisms. With further increase in \(\tau_s\), the velocity, temperature, and concentration fields approach a steady saturated state, demonstrating that nonlinear effects arrest the growth of the instability and lead to a stable finite-amplitude convection pattern.

Figure~\ref{fig:10v0.5k0.66IcRR.pdf}(a) shows the basic-state cell concentration profile for the rigid-rigid boundary configuration with $K_c=0.66$, $V=10$, and $\kappa=0.5$. The corresponding neutral stability curves are presented in Figure~\ref{fig:10v0.5k0.66IcRR.pdf}(b), where the critical bioconvection Rayleigh number is plotted against the wavenumber for different values of the thermal Rayleigh number $R_{t0}$. For $R_{t0}=0$, the stationary and oscillatory branches intersect at a wavenumber smaller than the critical wavenumber ($\lambda<\lambda_c$). Consequently, the stationary mode remains the most unstable mode at the onset of convection. A similar behavior is observed for all values of $R_{t0}$, where the oscillatory branch always merges with the stationary branch before $\lambda_c$, and the instability is therefore governed by the stationary mode. As the thermal Rayleigh number decreases ($R_{t0}<0$), the critical bioconvection Rayleigh number increases, indicating that thermal stratification stabilizes the system. In contrast, increasing the thermal Rayleigh number ($R_{t0}>0$) lowers the critical bioconvection Rayleigh number, thereby promoting the onset of convection. Furthermore, when the thermal Rayleigh number reaches the classical critical Rayleigh number for Rayleigh-B\'{e}nard convection with rigid-rigid boundaries, $R_{t0}=1707.77$, the critical bioconvection Rayleigh number becomes zero. At this point, the influence of the bioconvection parameters vanishes, and the onset of convection is governed entirely by thermal buoyancy, corresponding to the classical Rayleigh-B\'{e}nard convection limit.

\section{Conclusion}
\label{sec8}

A weakly nonlinear stability analysis has been performed to examine the influence of thermal buoyancy on phototactic bioconvection in a suspension of swimming microorganisms subjected to bottom heating or cooling. The equilibrium state, together with the corresponding linear stability, adjoint, and second-order resolvent problems, was solved numerically to derive the Stuart–Landau amplitude equation.  This formulation provides a unified framework for investigating both stationary and oscillatory nonlinear states under thermal forcing. The linear stability analysis shows that thermal buoyancy significantly alters both the onset and the mode of instability. Cooling from below increases the critical bioconvection Rayleigh number and delays the onset of convection, whereas heating from below progressively decreases the instability threshold. As the thermal Rayleigh number approaches the classical Rayleigh–B\`{e}nard critical value, the critical bioconvection Rayleigh number tends to zero, indicating a gradual transition from phototaxis-dominated to thermally driven convection.
 The weakly nonlinear analysis revealed that the perturbation amplitude initially grows from a small perturbation and eventually saturates to a finite equilibrium state due to nonlinear effects. For stationary instability, the system undergoes a supercritical pitchfork bifurcation in which two stable finite-amplitude branches emerge continuously from the basic state once the critical condition is exceeded. The vector field and potential function analyses further confirmed the transition from a single stable equilibrium state to two symmetric stable states. Moreover, the equilibrium amplitude increases with the thermal Rayleigh number, indicating that heating intensifies finite-amplitude convection. For higher cell swimming speeds and higher optical depth, oscillatory instability was observed. In this regime, the system exhibits sustained periodic oscillations characterized by stable limit cycles in the phase plane. The trajectories evolve away from the unstable equilibrium state and approach a closed periodic orbit, confirming the occurrence of a supercritical Hopf bifurcation. The real and imaginary parts of the complex amplitude remain bounded and periodic in time, demonstrating nonlinear saturation of the oscillatory mode. For parameter regimes exhibiting stationary onset, heating modifies the instability threshold and finite-amplitude convection without altering the bifurcation type, which remains supercritical pitchfork. In contrast, for parameter regimes exhibiting oscillatory onset, increasing the thermal Rayleigh number eventually changes the primary instability from oscillatory to stationary, resulting in a transition from a supercritical Hopf bifurcation to a supercritical pitchfork bifurcation. The thermal contribution is zero in the absence of thermal forcing, assumes negative values under cooling from below, and increases monotonically with bottom heating, demonstrating the increasing role of thermal buoyancy in the nonlinear regime.
Future work may consider additional physical effects, including light scattering, porous media, anisotropic illumination, and oblique irradiation, to further investigate the nonlinear dynamics of phototactic suspensions.

\bibliography{aipsamp}


\end{document}